\documentclass[12pt,reqno]{article}
\pdfoutput=1
\usepackage[usenames,dvipsnames]{xcolor}

\usepackage{amsmath,mathtools,tikz,pgfplots}
\usepackage{mathrsfs}
\usepackage{amssymb,amsthm,mathtools,bbm}
\usepackage{enumitem,cite}
\usepackage[longnamesfirst]{natbib}
\usepackage{multirow,abstract}
\usepackage{framed,mdframed,cancel,setspace}
\usepackage[bottom,hang]{footmisc}
\usepackage{graphicx,tikz}
\usepackage[colorlinks=true,allcolors=NavyBlue,citecolor=NavyBlue]{hyperref}
\hypersetup{breaklinks=true}
\usetikzlibrary{calc,positioning}
\usepackage[font={small,it}]{caption}
\usepackage[noabbrev]{cleveref}
\usepackage[T1]{fontenc}

\usepackage{breakurl}

\newtheoremstyle{ans}{15pt}{20pt}{}{}{\bfseries}{}{ }{\thmname{#1}\thmnote{ #3}.}
\theoremstyle{ans}
\newtheorem*{soln}{Solution}

\newcounter{lemno}
\newcounter{app_lemno}
\newcounter{propno}
\newcounter{app_propno}
\newcounter{thmno}
\newcounter{clmno}
\newcounter{app_thmno}
\newcounter{defno}
\newcounter{factno}
\newcounter{corno}
\newcounter{assno}
\newcounter{reassno}
\newcounter{examplenum}
\newcounter{remarkno}
\newmdenv[backgroundcolor=blue!7,linewidth=0pt]{bBox}
\newmdenv[backgroundcolor=green!7,linewidth=0pt]{gBox}
\newmdenv[backgroundcolor=red!7,linewidth=0pt]{rBox}
\newmdenv[backgroundcolor=gray!7,linewidth=0pt]{grayBox}

\newtheoremstyle{statementStyle}
{12pt}
{12pt}
{}
{}
{}
{{\bf .}\;}
{0.25em}
{{\bf\thmname{#1}\thmnumber{ #2}}
\thmnote{ (#3)}} 
\makeatother

\theoremstyle{statementStyle}

\theoremstyle{plain}
\newtheorem{lemma}[lemno]{Lemma}

\newtheorem{theorem}[thmno]{Theorem}

\newtheorem{coro}[corno]{Corollary}
\newtheorem{exerciseT}[thmno]{Exercise}
\newtheorem{exampleT}[examplenum]{Example}
\newtheorem{defn}[defno]{Definition}

\Crefname{assumption}{Assumption}{Assumptions}
\newtheorem{proposition}[propno]{Proposition}
\Crefname{proposition}{Proposition}{Propositions}

\Crefname{remark}{Remark}{Remarks}

\newtheorem{claim}[clmno]{Claim}
\Crefname{claim}{Claim}{Claims}

\newtheoremstyle{nbstatementStyle}
{12pt}
{12pt}
{\it}
{}
{}
{{\bf .}\;}
{0.25em}
{{\bf\thmname{#1}}
\thmnote{ (#3)}} 
\makeatother

\theoremstyle{nbstatementStyle}

\newtheoremstyle{restatementStyle}
{12pt}
{12pt}
{}
{}
{\bfseries}
{{\bf .}\;}
{0.25em}
{\thmname{#1}\thmnumber{ #2'}
\thmnote{ (#3)}} 
\makeatother

\theoremstyle{restatementStyle}

\newtheorem{prob}{Problem}

\newtheorem{conj}{Conjecture}
\newtheorem{prope}{Property}

\newtheoremstyle{reexerciseStyle}
{12pt}
{12pt}
{}
{}
{\bfseries}
{{\bf .}\;}
{0.25em}
{\thmname{#1}\thmnumber{ #2}
\thmnote{ (#3)}} 
\makeatother

\newtheoremstyle{exerciseStyle}
{12pt}
{12pt}
{}
{}
{\bfseries}
{{\bf .}\;}
{0.25em}
{\thmname{#1}\thmnumber{ #2}
\thmnote{ (#3)}} 
\makeatother

\theoremstyle{exerciseStyle}

\newtheoremstyle{teach}{15pt}{\topsep}{}{}{\itshape}{}{ }{\thmname{#1}\thmnote{ #3}.}
\theoremstyle{teach}
\newtheorem*{prooflemma}{Proof of lemma}

\newtheoremstyle{rem}{5pt}{0pt}{\color{black}}{}{\bfseries}{}{ }{\thmname{#1}\thmnote{ #3}.}
\theoremstyle{rem}

\definecolor{darkgreen}{rgb}{0.0, 0.75, 0.2}

\makeatletter
\renewcommand*\env@matrix[1][*\c@MaxMatrixCols c]{%
  \hskip -\arraycolsep
  \let\@ifnextchar\new@ifnextchar
  \array{#1}}
\makeatother

\renewcommand{\a}{\alpha}

\renewcommand{\d}{\delta}

\newcommand{\dd}{\mathrm{d}}

\newcommand{\la}{\lambda}
\newcommand{\pp}{\partial}

\newcommand{\La}{\Lambda}

\newcommand{\R}{\mathbb{R}}

\newcommand{\s}{\sigma}

\newcommand{\x}{\mathbf{x}}

\newcommand{\BAR}{\overline}

\newcommand{\und}[1]{\underline{#1}}

\DeclarePairedDelimiter{\abs}{\lvert}{\rvert}

\newcommand{\AND}{\quad\text{and}\quad}

\renewcommand{\(}{\left(}
\renewcommand{\)}{\right)}

\newcommand{\eg}{e.g.}
\newcommand{\ie}{i.e.}

\DeclareMathOperator*{\argmax}{arg\,max}

\DeclareMathOperator{\E}{\mathbf{E}}

\def\calI{\mathcal{I}}

\def\calL{\mathcal{L}}

\def\calO{\mathcal{O}}

\usepackage{datetime}

\newdateformat{monthyeardate}{%
  \monthname[\THEMONTH] \THEYEAR}
\newdateformat{daymonthyeardate}{%
  \monthname[\THEMONTH] \THEDAY, \THEYEAR}

\title{\LARGE\hspace*{\fill}\\[1ex]\papertitle\\[2ex]}
\author{\large\name\\[1ex] \affiliation}
\date{\monthyeardate\today}

\emergencystretch=\maxdimen
\makeatletter
\def\@xfootnote[#1]{%
  \protected@xdef\@thefnmark{#1}%
  \@footnotemark\@footnotetext}
\makeatother

\makeatletter
\ifFN@para
\else
  \long\def\@makefntext#1{%
    \ifFN@hangfoot
      \bgroup
      \setbox\@tempboxa\hbox{%
        \ifdim\footnotemargin>0pt
          \hb@xt@\footnotemargin{\@makefnmark\hss}%
        \else
          \@makefnmark\hskip-\footnotemargin      
        \fi
      }%
      \leftmargin\wd\@tempboxa
      \rightmargin\z@
      \linewidth \columnwidth
      \advance \linewidth -\leftmargin
      \parshape \@ne \leftmargin \linewidth
      \footnotesize
      \@setpar{{\@@par}}%
      \leavevmode
      \llap{\box\@tempboxa}%
      \parskip\hangfootparskip\relax
      \parindent\hangfootparindent\relax
    \else
      \parindent1em
      \noindent
      \ifdim\footnotemargin>\z@
        \hb@xt@ \footnotemargin{\hss\@makefnmark}%
      \else
        \ifdim\footnotemargin=\z@
          \llap{\@makefnmark}%
        \else
          \llap{\hb@xt@ -\footnotemargin{\@makefnmark\hss}}%
        \fi
      \fi
    \fi
    \footnotelayout#1%
    \ifFN@hangfoot
      \par\egroup
    \fi
  }
\fi
\makeatother

\makeatletter
\def\@endtheorem{\endtrivlist}
\makeatother

\newcommand{\citepos}[1]{\citeauthor{#1}'s (\citeyear{#1})}

\usepackage{booktabs,array,tablefootnote,subcaption,soul} 
\usepackage[titletoc]{appendix}
\newcolumntype{C}[1]{>{\centering\arraybackslash}m{#1}}

\makeatletter
\newcommand{\vast}{\bBigg@{4}}
\newcommand{\Vast}{\bBigg@{5}}
\makeatother

\DeclareMathOperator*{\LF}{LF}

\DeclareMathOperator{\co}{\mathbf{co}}

\definecolor{azure(colorwheel)}{rgb}{0.0, 0.5, 1.0}
\definecolor{brandeisblue}{rgb}{0.0, 0.44, 1.0}
\definecolor{ceruleanblue}{rgb}{0.16, 0.32, 0.75}
\definecolor{airforceblue}{rgb}{0.36, 0.54, 0.66}
\definecolor{bleudefrance}{rgb}{0.19, 0.55, 0.91}
\definecolor{darkspringgreen}{rgb}{0.09, 0.45, 0.27}
\definecolor{rulecolor}{rgb}{0.13, 0.35, 0.5}

\usepgfplotslibrary{fillbetween,decorations.softclip}
\usetikzlibrary{arrows.meta}
\usetikzlibrary{patterns,calc}

\newcommand{\papertitle}{{\bf Optimal In-Kind Redistribution}\footnote[$*$]{\thanktext}
}

\newcommand{\name}{Zi Yang Kang\footnote[$\dag$]{\affiliationi}\qquad Mitchell Watt\footnote[$\ddag$]{\affiliationii}}

\newcommand{\affiliationi}{Department of Economics, University of Toronto; \href{mailto:zy.kang@utoronto.ca}{\tt zy.kang@utoronto.ca}.}
\newcommand{\affiliationii}{Department of Economics, Stanford University; \href{mailto:mwatt@stanford.edu}{\tt mwatt@stanford.edu}.}
\newcommand{\thanktext}{We are especially indebted to Laura Doval, Marcin P\k{e}ski, and Andy Skrzypacz for many illuminating discussions.  We thank Piotr Dworczak, Marina Halac, Rishabh Kirplani, Ilya Segal, Xianwen Shi, Philipp Strack, and Satoru Takahashi for helpful comments and suggestions.  In addition, Kang gratefully acknowledges the support of the Center of Mathematical Sciences and Applications at Harvard University; and Watt gratefully acknowledges the support of the Koret Fellowship, the Ric Weiland Graduate Fellowship, and the Gale and Steve Kohlhagen Fellowship in Economics at Stanford University.
}
\newcommand{\abstracttext}{
\noindent This paper develops a model of in-kind redistribution where consumers participate in either a private market or a government-designed program, but not both.  We characterize when a social planner, seeking to maximize weighted total surplus, can strictly improve upon the laissez-faire outcome.  We show that the optimal mechanism consists of three components: a public option, nonlinear subsidies, and laissez-faire consumption.  We quantify the resulting distortions and relate them to the correlation between consumer demand and welfare weights.  Our findings reveal that while private market access constrains the social planner's ability to redistribute, it also strengthens the rationale for non-market allocations.
\\[2ex]
{\em JEL classification}: C61, D47, D63, D82, H42 \\[1ex]
{\em Keywords}: in-kind transfers, redistribution, optimal mechanism design, public provision}
\newcommand{\inserttitle}{\begin{center}\hfill\\[0ex]
\LARGE\papertitle\\[3ex]
\large
\name\\[3ex]
{
\monthyeardate\today}
\\[4ex]
\end{center}}

\begin{document}
\onehalfspacing
\inserttitle
\begin{abstract}
\abstracttext
\end{abstract}
\onehalfspacing
\thispagestyle{empty}

\clearpage

\setcounter{page}{1}

\section{Introduction}
\label{sec:introduction}

Governments often redistribute in kind by intervening in markets.  Two common instruments for in-kind redistribution are subsidy programs, which allow consumers to purchase different versions of the good at subsidized prices, and direct provision, which offers consumers access to a standard baseline version---sometimes referred to as the ``public option.''  For example, in the market for low-income housing, housing authorities frequently provide a combination of housing assistance programs, such as the Low-Income Housing Tax Credit in the United States, and public housing developments.  When and how should each of these instruments be used optimally?

In this paper, we answer these questions using a mechanism design approach.  We begin by considering a perfectly competitive market where consumers can purchase different quality levels of a good at cost.  We then introduce a social planner who seeks to redistribute by offering an alternative price schedule---such as by subsidizing some quality levels and/or making a baseline quality level freely available.  Following a recent literature on redistributive mechanism design, we model the social planner's redistribution objective by assigning heterogeneous welfare weights to consumers.  Each consumer's welfare weight represents the social value of giving him one unit of money; we interpret those with high welfare weights as ``poor'' and those with low welfare weights as ``rich.''  
In addition, to focus on in-kind redistribution, we restrict the social planner’s ability to redistribute in cash by disallowing lump-sum transfers to consumers.  Instead, we model the social planner's opportunity cost for money (\eg, due to other redistribution programs, including lump-sum transfers outside of the mechanism) by assigning a welfare weight to profit.

A key novel feature of our paper is the social planner's limited control over the entire market due to consumers' ability to access the private market.  On one hand, the social planner faces a screening problem: she seeks to distort the quality consumption of richer consumers and redirect surplus to poorer consumers.  On the other hand, unlike standard screening problems, the social planner faces additional participation constraints that arise because consumers have access to a private market, which restricts the social planner's ability to distort quality consumption and redirect surplus.  These participation constraints reflect the limited ability of housing authorities (and other public assistance programs, which are often run by local governments) to control prices (\eg, through nonlinear taxes) in the private market.

Our first main result quantifies {\em when} there is scope for in-kind redistribution.  As we show, the social planner optimally intervenes if and only if the welfare weight of the poorest consumer exceeds the welfare weight of profit.  One direction of this result is intuitive: if even the poorest consumer's welfare weight does not exceed that of profit, the social planner cannot improve upon the laissez-faire outcome and would prefer to use other programs or lump-sum transfers outside the mechanism, as captured by her high opportunity cost for money.  The reverse direction is less obvious: if the welfare weight of the poorest consumer exceeds profit, then we demonstrate that the social planner can always strictly improve on the laissez-faire outcome by distorting the quality consumption of some consumers and redirecting surplus to poorer consumers.

However, the ability of consumers to access a private market makes it more complicated to quantify the optimal distortion.  While the scope for in-kind redistribution depends on the welfare weight of the {\em poorest} consumer, the set of consumers for whom the social planner optimally distorts consumption---as we show---depends on the welfare weight of the {\em average} consumer.  On one hand, if the welfare weight of the average consumer exceeds that of profit, then only the participation constraint of the richest consumer can potentially bind.  In this case, the social planner optimally distorts the quality consumption of the richest consumers and redirects surplus to the poorest consumers.  On the other hand, if the welfare weight of the average consumer does not exceed that of profit, then the participation constraints bind for a positive measure of the richest consumers.  In this case, the social planner optimally distorts only the quality consumption of consumers in the middle, rather than the richest consumers.

The relationship between each consumer's welfare weight and willingness to pay introduces an additional layer of complexity.  Specifically, we consider two benchmark cases: when welfare weight is {\em negatively} correlated with willingness to pay, and when it is {\em positively} correlated.  A negative correlation implies that consumers who consume higher quality levels (\eg, bigger apartments) tend to be richer.  When some participation constraints bind, this leads to {\em downward} distortions for richer consumers and {\em upward} distortions for poorer consumers in the set of consumers who benefit from the optimal redistribution program.  By contrast, a positive correlation implies that consumers who consume higher quality levels (\eg, more childcare) tend to be poorer.  This leads to {\em upward} distortions for all consumers whose quality consumption the social planner distorts.

\begin{figure}[ht!]
\centering
\begin{subfigure}[t]{\textwidth}
\centering
\begin{tikzpicture}[scale=1.4]
\draw[-latex,very thick] (-5,0) -- (5.2,0) ; 
\node[below right] at (5, 0.) {richer consumers};

\draw[line width=5pt,darkspringgreen] (-4.5,0) -- (-3.5,0) node[above] {public option} -- (-2.5,0);
\draw[line width=5pt,orange] (-2.5,0) -- (-0.25,0) node[above] {subsidy program} -- (2,0);
\draw[line width=5pt,magenta] (2,0) -- (3.25,0) node[above] {private market} -- (4.5,0);
\draw[thick] (-4.5, 0.15) -- (-4.5, -0.15);
\draw[thick] (-2.5, 0.15) -- (-2.5, -0.15);
\draw[thick] (2, 0.15) -- (2, -0.15);
\draw[thick] (4.5, 0.15) -- (4.5, -0.15);

\end{tikzpicture}
\caption{Negative correlation between welfare weight and willingness to pay}
\label{fig:summary_negative}
\end{subfigure}
\vspace{15pt}

\begin{subfigure}[t]{\textwidth}
\centering
\begin{tikzpicture}[scale=1.4]
\draw[-latex,very thick] (-5,0) -- (5.2,0) ; 
\node[below right] at (5, 0.) {poorer consumers};

\draw[line width=5pt,magenta] (-4.5,0) -- (-1.5,0) node[above] {{\color{darkspringgreen}public option} {\color{black}or} {\color{magenta}private market}} -- (1.5,0);
\draw[line width=5pt,darkspringgreen,dashed] (-4.5,0) -- (1.5,0);
\draw[line width=5pt,orange] (1.5,0) -- (3.,0) node[above] {subsidy program} -- (4.5,0);
\draw[thick] (-4.5, 0.15) -- (-4.5, -0.15);
\draw[thick] (1.5, 0.15) -- (1.5, -0.15);
\draw[thick] (4.5, 0.15) -- (4.5, -0.15);

\end{tikzpicture}
\caption{Positive correlation between welfare weight and willingness to pay}
\label{fig:summary_positive}
\end{subfigure}\vspace{5pt}

\caption{Graphical depictions of the optimal in-kind redistribution program.}\label{fig:summary}
\end{figure}
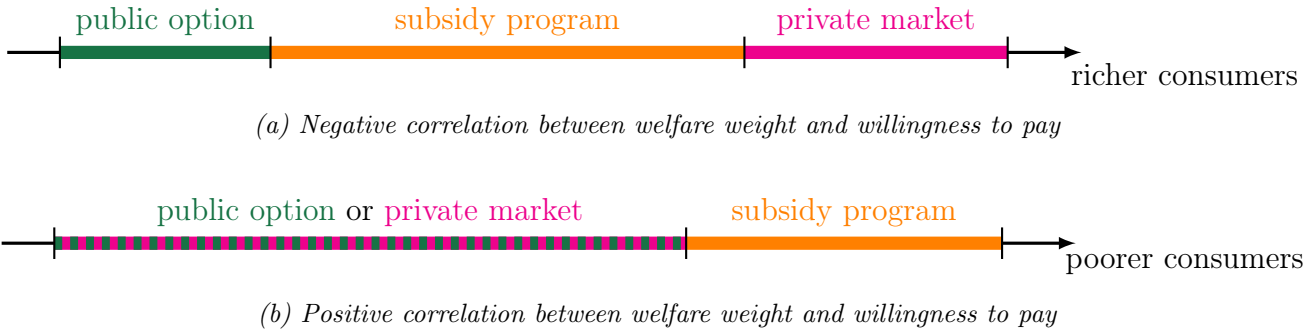

Despite these challenges, our second main result characterizes {\em how} the social planner should optimally redistribute in kind.  As we show, the optimal redistribution program generally comprises three components: a public option, a nonlinear subsidy program, and private market consumption.  How these components are structured depends on the correlation between each consumer's welfare weight and willingness to pay.  In the case of negative correlation, the poorest consumers receive a free public option, consumers in the middle benefit from a nonlinear subsidy, and the richest consumers consume in the private market, as illustrated in \Cref{fig:summary_negative}.  By contrast, in the case of positive correlation, the richest consumers consume either the public option or in the private market, and the poorest consumers benefit from a nonlinear subsidy, as illustrated in \Cref{fig:summary_positive}.

In the case of negative correlation between welfare weight and willingness to pay, our second result demonstrates how private market access can broaden the justification for using non-market allocations.  We compare our optimal redistribution program to a counterfactual scenario in which the social planner can control the entire market.  In that scenario, the social planner employs a public option only when the welfare weight of the average consumer exceeds that of profit.  In our problem, a public option remains optimal under that condition; however, when welfare weight is negatively correlated with willingness to pay, a public option may also be optimal even when that condition fails to hold.  This is because a public option may help to relax participation constraints that would otherwise bind for poorer consumers, resulting in broadened justification for its use. 

In the case of positive correlation between welfare weight and willingness to pay, our second result shows how the social planner does not benefit from preventing subsidized consumers from topping up their consumption in the private market.  Unlike affordable housing programs, programs such as childcare and disability care often do not explicitly prohibit consumers from supplementing their subsidized consumption in the program with additional consumption in the private market.  As we show in a companion paper (\citealp{kangwatt24a}), the ability of consumers to top up their consumption in the private market leads to different participation constraints and different optimal mechanisms in general.  However, we show in this paper that, when welfare weight is positively correlated with willingness to pay, the ability of consumers to top up their consumption in the private market does not lead to a different optimal mechanism.  Positive correlation ensures better self-targeting for those with higher welfare weights, and so the optimal mechanism is sufficiently generous such that no subsidized consumer tops up his consumption in the private market. 

Our characterization of the optimal in-kind redistribution program in this paper builds on the Lagrangian approach and generalized ironing methods in mechanism design.  The participation constraints that arise from consumers' ability to access a private market yield a mechanism design problem with type-dependent outside options.\footnote{\cite{jullien00} uses optimal control to characterize the solution to such a problem when the optimal mechanism is fully separating (\ie, it induces different consumers to consume different quality levels).  By contrast, optimal mechanisms in our paper are generally not fully separating (as shown in \Cref{fig:summary}).  While \citeauthor{jullien00} also characterizes necessary conditions that a non-fully separating optimal mechanism must satisfy, we explicitly characterize the optimal mechanism by guessing optimal Lagrange multipliers and verifying that they satisfy these conditions.}  As we show, this gives rise mathematically to a convex program with majorization constraints.\footnote{Recent advances (\eg, \citealp{kleineretal21}) have used extreme point methods to characterize the solutions of linear---and, more generally, concave---programs with majorization constraints.  However, extreme point methods do not apply in our setting as our program is convex; hence, our solution is at an interior point.}  We solve this problem by guessing Lagrange multipliers {\em \`a la} \cite{amadorbagwell13} and showing that they are optimal using generalized ironing methods developed by \cite{toikka11}. 

Our paper contributes to a growing literature that analyzes redistribution through the lens of mechanism design.  While much of this literature (\eg, \citealp{cheetal13}; \citealp{condorelli13}; \citealp{dworczaketal21}; \citealp{akbarpouretal24}; \citealp{paistrack24}) assumes that the social planner can control the entire market, we consider a social planner who cannot control the private market.  This makes our paper closer to \citepos{kang23}, which studies the equilibrium effects of a social planner's direct provision program on the private market.  Whereas \citeauthor{kang23} focuses on a benchmark where the social planner is more inefficient than the private market (as modeled by the restriction that she can produce only one quality level), we study an alternative benchmark where the social planner is just as efficient as the private market.  In addition, rather than equilibrium effects, our paper focuses on the problem of participation constraints (\ie, type-dependent outside options) that arises from the social planner's partial control of the market. 
As we explain later in the paper, our results can nevertheless also be extended to accommodate equilibrium effects using a similar approach to \citeauthor{kang23}.

Our paper focuses on a setting where consumers must choose between consuming in the private market or participating in the social planner's redistribution program.  In our view, this assumption is reasonable in settings like public housing, where it might be easier for public housing authorities and/or local government agencies to prevent consumers from renting both affordable housing and private apartments at the same time.  However, we recognize that this assumption might be less appropriate for public assistance programs such as food stamps and public transit subsidies.  These programs often allow---and sometimes expect---consumers to supplement subsidized goods with additional purchases in the private market.  While we show in this paper that consumers' ability to top up consumption in the private market does not change the optimal mechanism when welfare weight is positively correlated with willingness to pay, we examine this problem in full generality in a companion paper (\citealp{kangwatt24a}).  There, we show that the social planner intervenes under different conditions with different optimal mechanisms in general due to additional implied constraints on the marginal prices that consumers face.  For example, we show that the social planner uses a different sufficient statistic---related to a {\em conditional average} of consumers' welfare weights, rather than their maximum---in order to determine when she should optimally intervene.  From a methodological perspective, the ability of consumers to top up consumption in the private market results in a convex program with constraints akin to first-order stochastic dominance rather than majorization; consequently, the tools developed in both papers to characterize optimal mechanisms are very different.

Finally, our paper also complements a large literature in public finance on the public provision of private goods.  Whereas this literature (\eg, \citealp{nicholszeckhauser82}; \citealp{blackorbydonaldson88}; \citealp{besleycoate91}; \citealp{gahvarimattos07}) typically focuses on in-kind redistribution on the extensive margin---that is, whether or not to provide a fixed quality level of the good at a given price---our mechanism-design approach allows us to study the intensive margin as well, where the social planner provides a continuum of quality levels at different prices.

The remainder of this paper is organized as follows.  We begin by developing a model of in-kind redistribution in \Cref{sec:model}.  We then present our main results: \Cref{sec:scope} quantifies when there is scope for in-kind redistribution, while \Cref{sec:optimal} characterizes the optimal redistribution program.  We use our main results to derive economic implications in \Cref{sec:implications} and discuss broader takeaways in \Cref{sec:discussion}.  \Cref{sec:conclusion} concludes.  We provide a detailed derivation of our main results in \ref{app:proof_main} and defer all other proofs to \ref{app:additional_proofs}.

\section{Model}
\label{sec:model}

In this section, we develop a model of in-kind redistribution.  To this end, we begin with a standard model of a private market and describe the laissez-faire equilibrium.  We then formulate the social planner's problem of optimal in-kind redistribution via mechanism design.

\subsection{Setup}

There is a unit mass of risk-neutral consumers who each demands a single unit of an indivisible good.  The good is available at different quality levels, denoted by $q\in[0,A]$, up to a maximum quality level $A$.  The good is supplied competitively by producers in a private market who face a constant marginal cost of $c>0$ for each unit of quality.  

While consumers have quasilinear preferences over money, they differ in their preferences over quality.  In particular, each consumer's preferences over quality are captured by a type $\theta\in[\und\theta,\BAR\theta]\subset\R_{++}$ that determines the utility $\theta v(q)$ that he derives from consuming a good of quality $q$.  We assume that $v:[0,A]\to\R_+$ is twice continuously differentiable such that $v'>0$ and $v''<0$ (\ie, $v$ is increasing and strictly concave); these assumptions are standard and capture consumers' diminishing marginal utility for quality.  For notational simplicity, we extend the domain of $(v')^{-1}$ to $\R$ by setting $(v')^{-1}(z)=0$ for $z\geq v'(0)$ and $(v')^{-1}(z)=A$ for $z\leq v'(A)$.  This allows individual demand curves to be written as
\[D(p,\theta) = (v')^{-1}\(\frac{p}{\theta}\).\]
Each consumer is privately informed about his type, which is drawn from an absolutely continuous cumulative distribution function $F$ with a positive density function $f$ on $[\und\theta,\BAR\theta]$.

\subsection{Laissez-Faire Equilibrium}

Next, we describe the laissez-faire equilibrium in the private market.  As the market is perfectly competitive, the price of a good with quality $q$ is equal to its cost, $cq$, due to the assumption of constant marginal cost.  Thus, each consumer solves the utility maximization problem:
\[U^{\LF}(\theta)\coloneq\max_{q\in[0,A]}\left[\theta v(q)-cq\right].\]
Thus, $U^{\LF}(\theta)$ denotes the laissez-faire utility that a consumer with type $\theta$ receives by consuming in the private market.  We further let $q^{\LF}(\theta)$ denote the laissez-faire quality level of the good (\ie, the quality level that the consumer would choose in the private market); because $v$ is strictly concave, $q^{\LF}(\theta)$ is uniquely defined for each $\theta$.

We remark that, while the assumption of constant marginal cost is strong, it can be relaxed.  More generally, the aggregate supply curve for quality might slope upwards, which means that any intervention might change the equilibrium price in the market.  As \cite{kang23} shows, equilibrium effects can lead to an alternate redistribution channel in the form of pecuniary externalities, in which case the optimal redistribution program must also balance between the direct effect of the social planner’s screening problem and the indirect effect of pecuniary externalities.  In this paper, we focus on the problem of participation constraints that arise from the social planner's partial control of the market, because we think that type-dependent outside options are realistic and pose a technical challenge independently of equilibrium effects.  However, our results can be extended to allow for equilibrium effects using an approach similar to that taken by \citeauthor{kang23}.

\subsection{Social Planner's Problem}

We now consider the problem faced by a social planner who wishes to redistribute by designing an alternative price schedule for different quality levels of the good.  Unlike the private market's price schedule, which is linear because the private market is perfectly competitive, the social planner's price schedule might potentially be nonlinear.  Each consumer can choose to consume from either the social planner's price schedule or the private market's, but not both.  In particular, as is the case in our motivating example of public housing in \Cref{sec:introduction}, a consumer cannot ``top up'' his consumption in the private market if he has already chosen to consume from the social planner's price schedule.
\footnote{This assumption is natural in the context of public housing, where it might be easier for government agencies to prevent consumers from consuming both public housing and private housing at the same time.  However, there are other settings where consumers {\em can} top up their consumption in the private market, such as in the context of food vouchers, where consumers who receive SNAP benefits can supplement their consumption of food and groceries by paying in cash.  The ability of consumers to top up changes the social planner's problem significantly, and we study this setting in a companion paper \citep{kangwatt24a}.}

The social planner shares the same production technology as producers in the private market.  In particular, the social planner faces the same marginal cost of $c$ for each unit of quality.  We interpret this as a setting in which the social planner is equally efficient as private producers.\footnote{\cite{kang23} analyzes a different benchmark in which the social planner is less efficient than private producers.}  For example, the social planner might be able to costlessly contract with producers in the private market to supply affordable housing.

Because the social planner is equally efficient as private producers, we can reformulate the social planner's problem as one of choosing {\em total} consumption, defined as the sum of public consumption (from the social planner's price schedule) and private consumption (from the private market's price schedule).  That is, the social planner chooses a direct mechanism $(q,t)$, which consists of:
\begin{enumerate}[label={\em(\roman*)}]
\item an allocation function $q:[\und\theta,\BAR\theta]\to[0,A]$, so that $q(\theta)$ is the quality level of the good that a consumer with type $\theta$ consumes, be it from the social planner's price schedule or the private market's; and
\item a payment function $t:[\und\theta,\BAR\theta]\to\R$, so that $t(\theta)$ is the payment that a consumer with type $\theta$ makes, be it to the social planner or the private market.
\end{enumerate}

The key assumption enabling this reformulation is that the social planner shares the same production technology as producers in the private market.  Indeed, this assumption implies that total consumption is sufficient for determining total surplus; otherwise, if production technologies were asymmetric between the social planner and the private market, total surplus would depend on the breakdown of total consumption between the social planner and the private market.  This reformulation also implicitly assumes that the social planner chooses a deterministic mechanism; as we show in \ref{app:proof_main}, this assumption entails no loss of generality.  

We now describe three feasibility constraints on mechanisms that the social planner faces.

First, by the revelation principle \citep{myerson81}, it is without loss of generality for the social planner to restrict attention to incentive-compatible mechanisms so that consumers report their types truthfully:
\begin{equation}\label{eq:IC}
    \theta\in\argmax_{\theta'\in[\und\theta,\BAR\theta]}\left[\theta v(q(\theta'))-t(\theta')\right]\qquad\text{for any }\theta\in[\und\theta,\BAR\theta].\tag{IC}
\end{equation}

Second, unlike mechanism design problems without a private market, the ability of consumers to consume in a private market imposes type-dependent individual rationality constraints on the mechanism.  Because each consumer always has the option of consuming in the private market, the utility that he derives from the mechanism (which describes his total consumption) must be no less than the utility that he derives from his private market consumption:\footnote{Although our analysis of these constraints stem from consumers' ability to alternatively consume in the private market, it is worth noting that identical constraints might arise in a mechanism design problem where the social planner can control the entire market, but is subject to the constraint that the mechanism she designs must lead to a weak Pareto improvement over the laissez-faire outcome.}
\begin{equation}\label{eq:IR}
    \theta v(q(\theta)) - t(\theta) \geq U^{\LF}(\theta)\qquad\text{for any }\theta\in[\und\theta,\BAR\theta].\tag{IR}
\end{equation}

Third, we focus on mechanisms for in-kind redistribution by restricting the ability of the social planner to redistribute through lump-sum cash transfers within the mechanism.  In particular, the social planner can choose only nonnegative payment functions: 
\begin{equation}\label{eq:LS}
t(\theta)\geq 0\qquad\text{for any }\theta\in[\und\theta,\BAR\theta].\tag{LS}
\end{equation}  
As such, while the social planner can give away goods for free (\ie, setting a price of zero for some quality levels), she cannot give consumers money.  However, we do allow for cash transfers outside of the mechanism, which we model through the social planner's preferences over the monetary surplus (\ie, profit) that the mechanism runs, as we now describe.

The social planner maximizes total weighted surplus, which consists of consumer surplus and total profit (\ie, total revenue minus total cost):
\begin{enumerate}[label={\em(\roman*)}]
    \item {\bf Consumer surplus.}  The social planner assigns a social welfare weight of $\omega(\theta)$ to a consumer with type $\theta$.  This represents the social value that arises from giving that consumer one unit of money.\footnote{\cite{dworczaketal21} provide a microfoundation for this interpretation by showing that $\theta$ can be thought of as the marginal rate of substitution between quality and money.}  Throughout, we assume that $\omega:[\und\theta,\BAR\theta]\to\R_+$ is continuous.  While different markets may entail different correlations between welfare weight and willingness to pay, we focus on two benchmarks in this paper: negative correlation and positive correlation.  

In the case of negative correlation, we assume that $\omega$ is decreasing: consumers with higher value $\theta$ for quality tend to be richer and/or more privileged, and so the social planner assigns a lower welfare weight $\omega(\theta)$ to these consumers.  We expect this to hold in markets for housing and education.

In the case of positive correlation, we assume that $\omega$ is increasing: consumers with higher value $\theta$ for quality tend to be poorer and/or more disadvantaged, and so the social planner assigns a higher welfare weight $\omega(\theta)$ to these consumers.  We expect this to hold in markets for childcare, disability care, and inferior goods. 

    \item {\bf Total profit.}  The social planner assigns a social welfare weight of $\a\in\R_+$ to total profit, which is the difference between the total revenue generated by the mechanism and its total cost.   This represents the social planner's opportunity cost of budget.  For example, we would expect $\a$ to be high if the social planner has other redistribution programs that compete for the same budget, such as cash transfers that happen outside of the mechanism.  By contrast, we would expect $\a$ to be low if the social planner faces significant political constraints\footnote{For instance, \cite{liscowpershing22} show that the general population in the U.S.~largely prefers in-kind redistribution to cash transfers.} that limit her ability to redistribute outside of the mechanism.
\end{enumerate}

In summary, the social planner chooses a feasible mechanism---satisfying the \eqref{eq:IC}, \eqref{eq:IR}, and \eqref{eq:LS} constraints---to maximize total weighted surplus.  Consequently, the social planner's problem can be formally written as:
\begin{align}
    \max_{(q,t)}&\int_{\und\theta}^{\BAR\theta}\bigg[\omega(\theta)\underbrace{\left[\theta v(q(\theta))-t(\theta)\right]}_{\text{consumer surplus}} + \a\underbrace{\left[t(\theta)-cq(\theta)\right]}_{\text{total profit}}\bigg]\ \dd F(\theta)\tag{OBJ}\label{eq:OBJ}\\
    \text{s.t. }&(q,t) \text{ satisfies \eqref{eq:IC}, \eqref{eq:IR}, and \eqref{eq:LS}.}\tag*{}
\end{align}
Notice that the laissez-faire mechanism, $(q^{\LF},cq^{\LF})$, is feasible; hence the set of feasible mechanisms is always nonempty.  In \ref{app:proof_main}, we formally show that there is generally a unique solution to the social planner's problem.\footnote{More precisely, we show in \ref{app:proof_main} that there is a unique optimal allocation function that solves the social planner's problem, and the optimal payment function is uniquely pinned down when $\E[\omega]\ne\a$.  However, when $\E[\omega]=\a$, then uniform cash transfers from the social planner to all consumers are welfare-neutral; hence, there can be multiple optimal payment functions.  Our subsequent analysis focuses on the optimal mechanism with the pointwise smallest payment function when $\E[\omega]=\a$ (and the unique optimal mechanism when $\E[\omega]\ne\a$).}  Below, we refer to that solution as the optimal mechanism and denote it by $(q^*,t^*)$.

\section{Scope of Optimal Intervention}\label{sec:scope}

In this section, we present our first main result: a characterization of when the social planner can strictly improve on the laissez-faire outcome. 

\begin{theorem}[scope of optimal intervention]\label{thm:scope}
    The optimal mechanism $(q^*,t^*)$ strictly improves on the laissez-faire outcome if and only if $\max \omega>\a$.
\end{theorem}

\Cref{thm:scope} states that the difference between the {\em maximum} consumer welfare weight and the welfare weight of profit is a sufficient statistic for determining if the social planner should intervene.  

One direction of this result is intuitive: if $\max\omega\leq\a$, then the opportunity cost of the social planner's budget is relatively high, so the the social planner optimally chooses not to intervene (\ie, she prefers to use other redistribution programs described by her opportunity cost of budget).  

The other direction of this result, however, is less obvious.  In the remainder of this section, we sketch the proof of \Cref{thm:scope} and discuss the role of participation constraints in this result.

\subsection{Proof Sketch of \texorpdfstring{\Cref{thm:scope}}{Theorem 1}}\label{sec:scope_proof}

To establish the reverse direction of \Cref{thm:scope}, we construct a strict improvement over the laissez-faire outcome when $\max\omega>\a$.  We start by considering the case of negative correlation (\ie, $\omega$ is decreasing), before discussing how this reasoning extends to more general cases.

\begin{figure}[t!]
\centering
\begin{tikzpicture}[scale=1]

\begin{scope}
    \clip (1,0) rectangle (4,6); 
    \fill[darkspringgreen, opacity=0.4] (1,0) -- plot[domain=1:4, smooth] (\x, {\x*\x/25+1.5}) -- 
          plot[domain=4:1, smooth] (\x, {\x*\x/25+0.5}) -- cycle;
\end{scope}

\begin{scope}
    \clip (4,0) rectangle (9,6); 
    \fill[orange, opacity=0.4] (1,0) -- plot[domain=4:9, smooth] (\x, {(\x-4)*0.32+2.14}) -- 
          plot[domain=9:4, smooth] (\x, {\x*\x/25+0.5}) -- cycle;
\end{scope}

\draw[scale=1, domain=1:11., densely dashed, variable=\x, line width=2.pt, color=black!50] plot ({\x}, {\x*\x/25+0.5});
\draw (7.5,2.5) node[right] {$\color{gray}U^{\LF}$};

\draw [-{latex[scale=1.2]}, line width=1.5pt, ceruleanblue] (0.9,0.5) -- (0.9,1) node [left] {$\Delta\s$} -- (0.9,1.5);

\draw[scale=1, domain=1:4., smooth, variable=\x, line width=2.pt, color=ceruleanblue] plot ({\x}, {\x*\x/25+1.5});
\draw[scale=1, domain=4:9., smooth, variable=\x, line width=2.pt, color=ceruleanblue] plot ({\x}, {(\x-4)*0.32+2.14});
\draw[scale=1, domain=9:11., smooth, variable=\x, line width=2.pt, color=ceruleanblue] plot ({\x}, {\x*\x/25+0.5});
\draw[line width=1.pt, color=gray] (4, 2.14) -- (4, 1.14);

\draw[line width=1.5pt, dashed, color=gray] (9,3.74) -- (9, 1.);
\draw[{latex[scale=1.2]}-{latex[scale=1.2]}, line width=1.5pt, ceruleanblue] (4,1.) -- (6.5,1.) node [above] {$\calO(\abs{\Delta\s}^{1/2})$} -- (9,1.);


\draw [-{latex[scale=1.2]}, line width=1.5pt] (0,0) node [left] {0} -- (0,0) -- (12,0) node [below] {$\theta$};
\draw [-{latex[scale=1.2]}, line width=1.5pt] (0, -.5) -- (0,0) -- (0,6) node [above] {utility, $U$};

\draw[line width=1.5pt,] (1,0.15) -- (1,-0.15) node[below] {$\und\theta$};
\draw[line width=1.5pt,] (11.,0.15) -- (11.,-0.15) node[below] {$\BAR\theta$};
\draw[line width=1.5pt,] (4.,0.15) -- (4.,-0.15) node[below] {$\kappa$};

\end{tikzpicture}
\caption{Proof sketch of \Cref{thm:scope} in the case of negative correlation.}
\label{fig:scope_proof}
\end{figure}
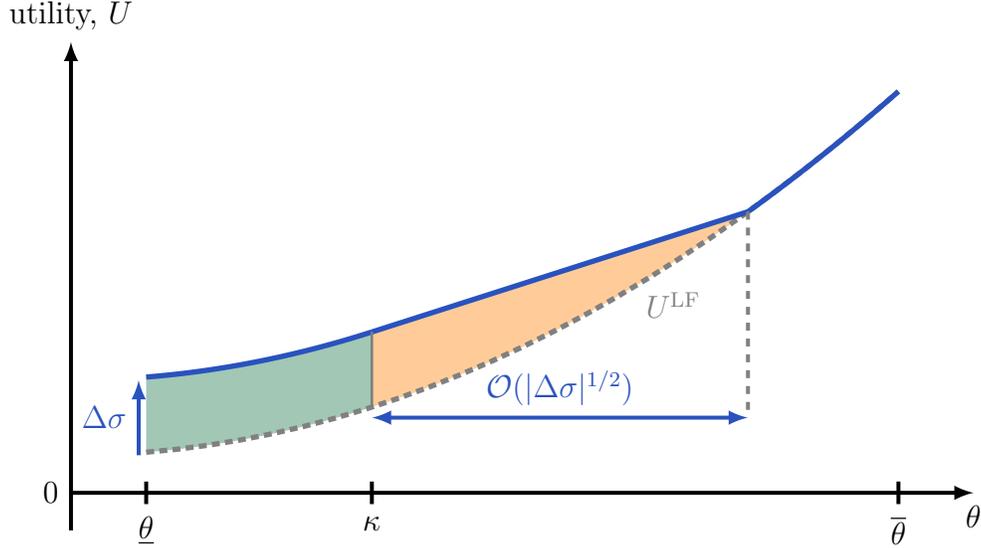

To this end, consider the utility that each consumer type receives, as shown in \Cref{fig:scope_proof}.  As a benchmark, the utility that each consumer type receives from the laissez-faire mechanism is shown as a gray dashed curve, $U^{\LF}$.  By the envelope theorem (cf.~\Cref{clm:IC} in \ref{app:proof_main}), the gradient of the utility curve must be nondecreasing in type for any incentive-compatible mechanism; hence, incentive-compatible mechanisms must result in convex utility curves.  In addition, individually rational mechanisms must result in utility curves that lie weakly above $U^{\LF}$.

Next, we fix a cutoff type $\kappa>\und\theta$ and suppose that the social planner gives a cash subsidy of magnitude $\Delta\s>0$ to consumer types in the interval $[\und\theta,\kappa]$.  This is shown by the blue utility curve in \Cref{fig:scope_proof}, which is a parallel upward shift of $U^{\LF}$ for consumer types in the interval $[\und\theta,\kappa]$.   Consumers with types just above $\kappa$ mimic the consumption of $\kappa$ in order to receive the cash subsidy $\Delta\s$: we thus extend the blue utility curve linearly from $\kappa$ until it intersects $U^{\LF}$.

The welfare effect of this cash subsidy is the sum of two components, represented by the green and orange areas in \Cref{fig:scope_proof}.  In the green area, since the cash subsidy does not distort consumption, the net impact on social welfare is exactly equal to $\Delta\s$, multiplied by the difference between the welfare weights of consumers and the opportunity cost of the social planner's budget:
\[\Delta\s\cdot\int_{\und\theta}^\kappa\left[\omega(\theta) - \a\right]\ \dd F(\theta).\]
In the orange area, the cash subsidy distorts consumption by no more than $\calO(\Delta\s)$; however, the measure of affected types is bounded from above by $\calO(\abs{\Delta\s}^{1/2})$. For these consumer types, the net impact on social welfare is thus bounded from above by $\calO(\abs{\Delta\s}^{3/2})$.  

By taking limits, we can see that this subsidy has a positive marginal impact in the case of negative correlation  when $\max\omega>\a$.  Indeed, the first-order term as $\Delta\s\to0$ and $\kappa\to\und\theta$ is equal to
\[\left[\omega(\und\theta) - \a\right]f(\und\theta)\cdot\Delta\s = \(\max\omega-\a\)f(\und\theta)\cdot\Delta\s>0.\]
This concludes our proof sketch of \Cref{thm:scope} in the case of negative correlation.

Although we have focused on the case of negative correlation, it is relatively straightforward to see that the above argument generalizes.  In the case of positive correlation, for example, the social planner can achieve a strict improvement over the laissez-faire outcome when $\max\omega>\a$ by offering a small cash subsidy to consumer types that are sufficiently close to $\BAR\theta$.  More generally, the social planner can always achieve a strict improvement by offering a small cash subsidy to consumer types that are sufficiently close to the consumer type with the highest welfare weight.

\subsection{Discussion of \texorpdfstring{\Cref{thm:scope}}{Theorem 1}}

We now explain the role that participation constraints play in \Cref{thm:scope} by comparing our result to a relaxed problem where the social planner faces the constraint that $\und U\geq0$ instead of the \eqref{eq:IR} constraints.  This relaxed problem can be interpreted as a ``full mechanism design'' problem where the social planner can control the entire market.  In this case, it is straightforward to show that the social planner can {\em always} strictly improve on the laissez-faire outcome (we provide details in \ref{app:proof_main}).  This is most easily seen when $\E[\omega]>\a$, which implies that the social planner's outside options are less efficient than the redistribution program.  However, even when $\E[\omega]\leq\a$, the social planner can always distort the quality consumption of richer consumers downwards by setting a higher marginal price for higher quality levels.  As such, the optimal mechanism taxes the quality consumption of richer consumers relative to the laissez-faire outcome.

Unlike this relaxed problem, participation constraints in \Cref{thm:scope} restrict the social planner's ability to tax the quality consumption of richer consumers.  For example, local housing authorities often have limited ability to tax rental prices in the private market.\footnote{Although our model introduces $c$ as the gross marginal cost of quality, we can accommodate limited forms of taxation (as we show in \citealp{kangwatt24a}) by interpreting $c$ as the marginal cost of quality net of taxes.}  Observe that:
\begin{enumerate}[label={\em(\roman*)}]
\item If $\E[\omega]>\a$, then participation constraints can bind for at most one type of consumer; thus, the logic of the relaxed problem extends to \Cref{thm:scope}.  Indeed, $\E[\omega]>\a$ implies that $\max\omega>\a$, so the social planner can strictly improve on the laissez-faire outcome.

\item If $\E[\omega]\leq\a$, then participation constraints generally bind for a positive measure of consumers; thus, the logic of the relaxed problem no longer extends to \Cref{thm:scope} due to the additional restrictions imposed by participation constraints.  In particular, the social planner {\em cannot} strictly improve on the laissez-faire outcome if and only if participation constraints bind for {\em all} consumers.
\end{enumerate}

In \ref{app:proof_main}, we provide an alternative proof of \Cref{thm:scope}, which reveals that the key determinant of when the participation constraints bind for all consumers depends on the sign of the correlation between welfare weight and willingness to pay.  Below, we consider the cases of negative correlation and positive correlation separately and show that the resulting conditions simplify to yield the sufficient statistic in \Cref{thm:scope}.
\begin{enumerate}[label={\em(\roman*)}]
\item {\bf Negative correlation.} If $\omega$ is decreasing, then participation constraints bind for all consumers if and only if $\E[\omega]\leq\a$ and
\begin{equation}\label{eq:scope_negative}
\int_{\und\theta}^{\theta}\left[\a-\omega(s)\right]\ \dd F(s) \geq 0\qquad\text{for every }\theta\in[\und\theta,\BAR\theta].
\end{equation}

To understand this condition, we show that the integral in \cref{eq:scope_negative} has the same sign of the distortion experienced by a consumer with type $\theta$, assuming that the participation constraint binds for the highest type, $\BAR\theta$.  Thus, this condition requires that the solution to the relaxed problem distorts the quality consumption of all consumers upwards.  Intuitively, if participation constraints do not bind for some consumers, then their consumption levels must be distorted upwards when the above condition holds.  However, given that the participation constraint binds for $\BAR\theta$, the envelope theorem implies that their resulting utility must be lower than that under the laissez-faire outcome, thereby contradicting the assumption that their participation constraints do not bind.

This condition further simplifies to give the sufficient statistic in \Cref{thm:scope}.  Given that $\omega$ is decreasing, it follows that $\theta\mapsto\int_{\und\theta}^\theta\left[\a-\omega(s)\right]\ \dd F(s)$ is quasiconvex.  Thus the condition described by \cref{eq:scope_negative} holds if and only if the gradient of $\theta\mapsto\int_{\und\theta}^\theta\left[\a-\omega(s)\right]\ \dd F(s)$ is nonnegative at $\theta=\und\theta$.  Equivalently, participation constraints bind for all consumers when $\a\geq \omega(\und\theta)=\max\omega$.  This condition implies that $\E[\omega]\leq\a$ and hence is a necessary and sufficient condition for participation constraints to bind for all consumers.

\item {\bf Positive correlation.} If $\omega$ is increasing, then participation constraints bind for all consumers if and only if $\E[\omega]\leq\a$ and
\begin{equation}\label{eq:scope_positive}
\int_{\theta}^{\BAR\theta}\left[\a-\omega(s)\right]\ \dd F(s) \geq 0\qquad\text{for every }\theta\in[\und\theta,\BAR\theta].
\end{equation}

The integral in \cref{eq:scope_positive} has the opposite sign of the distortion experienced by a consumer with type $\theta$, assuming that the participation constraint binds for the lowest type, $\und\theta$.  Thus, a symmetric argument to the one above for negative correlation---using the fact that $\theta\mapsto\int_{\theta}^{\BAR\theta}\left[\a-\omega(s)\right]\ \dd F(s)$ is quasiconvex when $\omega$ is increasing---conveys the intuition for why the condition in \eqref{eq:scope_positive} is necessary and sufficient for all participation constraints to bind.
\end{enumerate}

We conclude this section by remarking that \Cref{thm:scope} captures only the {\em marginal} incentives for the social planner to intervene.  In particular, despite the simplicity of the argument in \Cref{sec:scope_proof} to show a strict improvement over the laissez-faire outcome through a small, targeted cash subsidy, the optimal mechanism---as we describe below---is more complicated.

%
%

\section{Optimal Mechanisms}
\label{sec:optimal}

We now explain the second main result of this paper: a characterization of the social planner's optimal mechanism.  To this end, we present and interpret the optimal mechanisms under negative and positive correlation separately, and then provide  intuition for these characterizations.

Our characterization of the social planner's optimal mechanism builds on the following ironing operator (cf.~\citealp{myerson81}; \citealp{toikka11}).  For any generalized function $h$ with domain $[\und\theta,\BAR\theta]$, let its ironed version $\BAR{h}:[\und\theta,\BAR\theta]\to\R$ be the nondecreasing function defined by
\[\BAR h(\theta) \coloneq \left.-\frac{\dd}{\dd z}\(z\mapsto \co\int_{F^{-1}(z)}^{\BAR\theta} h(s)\ \dd F(s)\)\right|_{z=F(\theta)},\]
where $\co H$ is the pointwise smallest concave function that is no smaller than the given function $H:[\und\theta,\BAR\theta]\to\R$.  We denote the Dirac delta function at any point $a\in\R_+$ by $\d_a$.

\subsection{Optimal Mechanisms Under Negative Correlation}

We begin by characterizing the optimal mechanism when welfare weight is negatively correlated with willingness to pay.

\begin{theorem}[characterization of optimal mechanisms under negative correlation]\label{thm:optimal_negative}
Suppose that $\omega$ is decreasing.  For any $\mu\in\R$, define 
\[q_\mu(\theta) \coloneq D\(c,\BAR{H_\mu}(\theta)\),\quad\text{where }H_\mu(\theta)\coloneq \theta + \frac{\mu\und\theta\cdot \d_{\und\theta}(\theta)+\mu+\int_{\und\theta}^{\theta}\left[\a-\omega(s)\right]\ \dd F(s)}{\a f(\theta)}.\]
Moreover, denote $\mu_{\max}\coloneq -\min_{\theta\in[\und\theta,\BAR\theta]}\int_{\und\theta}^\theta\left[\a-\omega(s)\right]\ \dd F(s)$, and define $\theta_H:[0,\mu_{\max}]\to[\und\theta,\BAR\theta]$ by
\[\theta_H(\mu)\coloneq \begin{dcases}
\max\left\{\theta\in[\und\theta,\BAR\theta]:\int_{\und\theta}^{\theta}\left[\a-\omega(s)\right]\ \dd F(s)+\mu\leq0\right\}&\text{if }\E[\omega]\leq \a,\\
\BAR\theta&\text{if }\E[\omega]>\a.
\end{dcases}\]
Let
\[\mu^* \coloneq \min\left\{\mu\in[\(\E[\omega]-\a\)_+,\mu_{\max}]:\int_{\und\theta}^{\theta_H(\mu)}v(q_{\mu}(s))\ \dd s + \und\theta v(q_{\mu}(\und\theta))- U^{\LF}(\theta_H(\mu))\geq0\right\}.\]
Then the optimal allocation function is 
\[q^*(\theta) = \begin{dcases}
q^{\LF}(\theta) &\text{if }\theta_H(\mu^*)<\theta\leq\BAR\theta,\\
q_{\mu^*}(\theta) &\text{for }\und\theta\leq \theta\leq \theta_H(\mu^*).
\end{dcases}\]
\end{theorem}

To understand \Cref{thm:optimal_negative}, we begin by considering the case where the average welfare weight of consumers exceeds the welfare weight of profit, $\E[\omega]>\a$.  In this case, $\theta_H$ is constant and equal to $\BAR\theta$ for all $\mu\in[0,\mu_{\max}]$; so \Cref{thm:optimal_negative} states that the optimal allocation function is $q^*(\theta)=q_{\mu^*}(\theta)$ for all $\theta\in[\und\theta,\BAR\theta]$, where
\begin{align*}
\mu^*
&= \min\left\{\mu\in[\(\E[\omega]-\a\)_+,0\},\mu_{\max}]:\int_{\und\theta}^{\BAR\theta}v(q_\mu(s))\ \dd s + \und\theta v(q_\mu(\und\theta)) - U^{\LF}(\BAR\theta)\geq 0\right\}\\
&= \min\left\{\mu\in[\(\E[\omega]-\a\)_+,\mu_{\max}]:\int_{\und\theta}^{\BAR\theta}\left[v(q^{\LF}(s))-v(q_\mu(s))\right]\ \dd s \leq \und\theta \left[v(q_\mu(\und\theta)) - v(q^{\LF}(\und\theta))\right]\right\}.
\end{align*}
In the latter inequality, observe that the left-hand side, $\int_{\und\theta}^{\BAR\theta}\left[v(q^{\LF}(s))-v(q_\mu(s))\right]\ \dd s$, is decreasing in $\mu$, while the right-hand side, $\und\theta \left[v(q_\mu(\und\theta)) - v(q^{\LF}(\und\theta))\right]$, is increasing in $\mu$.  Note also that
\begin{align*}
\theta\leq H_{\mu_{\max}}(\theta)\implies \int_{\und\theta}^{\BAR\theta}\left[v(q^{\LF}(s))-v(q_{\mu_{\max}}(s))\right]\ \dd s
&\leq 0 \leq \und\theta\left[v(q_{\mu_{\max}}(\und\theta))-v(q^{\LF}(\und\theta))\right].
\end{align*}
Because terms in the inequality are continuous in $\mu$, this implies that $\mu^*$ is well-defined.

We now highlight three technical properties of the optimal mechanism and their corresponding economic interpretations when $\E[\omega]>\a$.  \Cref{fig:optimal_negative_allocation} shows an example of the optimal allocation function in this case.
\begin{enumerate}[label={\em(\roman*)}]
\item {\em $\mu^*>0$: the optimal redistribution program always includes a free public option.}

Observe that $\mu^*>0$ as $\mu^*\geq \(\E[\omega]-\a\)_+>0$ when $\E[\omega]>\a$.  This means that $H_{\mu^*}$ has an atom at $\und\theta$; hence, $\BAR{H_{\mu^*}}$---and, by extension, $q^*=q_{\mu^*}$---must be flat in a neighborhood of $\und\theta$.  

We interpret the resulting flat region of $q^*$ as a free public option.  Since $q^*$ is increasing, $q^*(\und\theta)$ is the lowest quality available in the market under the optimal redistribution program.  Consequently, $q^*(\und\theta)$ can be viewed as a ``baseline'' public option the social planner provides in the market.  Moreover, we show in the proof of \Cref{thm:optimal_negative} that $\mu^*$ is precisely the shadow cost of the no lump-sum transfer \eqref{eq:LS} constraint.  Given that the \eqref{eq:LS} constraint binds, the public option is free of charge.

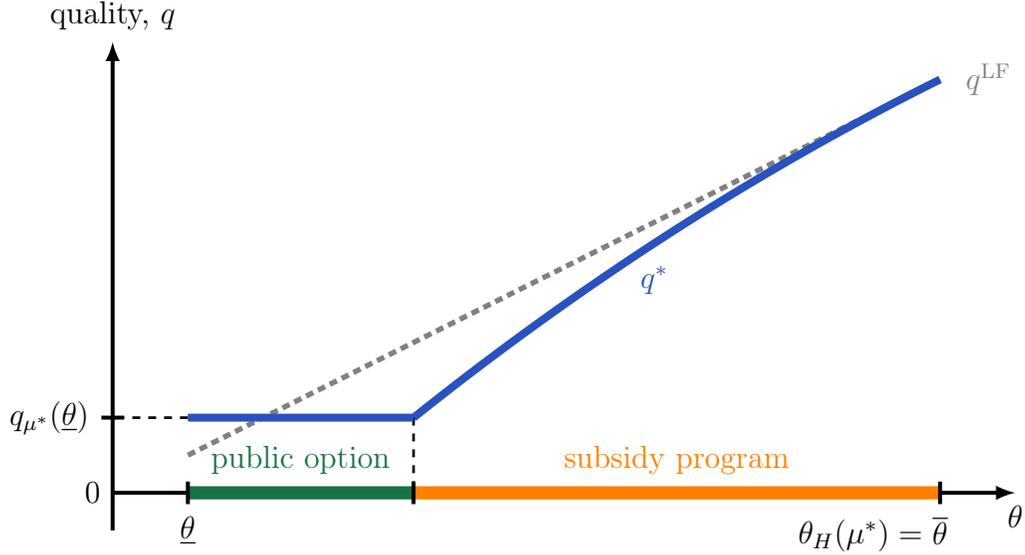
\begin{figure}[t!]
\begin{center}
\begin{tikzpicture}[scale=1.]

\draw[scale=1, domain=1:11., densely dashed, variable=\x, line width=2.pt, color=black!50] plot ({\x}, {0.5*\x});
\draw (11.2,5.5) node[right] {$\color{black!50}q^{\LF}$};

\draw (7.2,2.5) node[above] {$\color{ceruleanblue}q^*$};

\draw[scale=1, domain=4:11., smooth, variable=\x, line width=3pt, color=ceruleanblue] plot ({\x}, {0.0002324*\x^3 - 0.02669 * \x^2 + 1.0017*\x - 2.5946});
\draw[line width=1pt, dashed, color=black] (4,1) -- (4,0.15);

\draw[scale=1, domain=1:4., smooth, variable=\x, line width=3pt, color=ceruleanblue] plot ({\x}, {1});
\draw[line width=1pt, dashed, color=black] (1,1) -- (0.15,1);
\draw[line width=1.5pt, color=black] (0.15,1) -- (-0.15,1) node[left] {$q_{\mu^*}(\und\theta)$};


\draw (2.5,0.1) node [above] {{\color{darkspringgreen} public option}};

\draw (7.5,0.1) node [above] {{\color{orange} subsidy program}};

\draw [-{latex[scale=1.2]}, line width=1.5pt] (0,0) node [left] {0} -- (0,0) -- (12,0) node [below] {$\theta$};
\draw [-{latex[scale=1.2]}, line width=1.5pt] (0, -.5) -- (0,0) -- (0,6) node [above] {quality, $q$};

\draw[line width=5pt,darkspringgreen] (1,0) -- (4,0);
\draw[line width=5pt,orange] (4,0) -- (11,0);

\draw[line width=1.5pt,] (1,0.15) -- (1,-0.15) node[below] {$\und\theta$};
\draw[line width=1.5pt,] (11.,0.15) -- (11.,-0.15) node[below] {$\BAR\theta$};
\draw (10.,-0.15) node[below] {$\theta_H(\mu^*)=\vphantom{\BAR\theta}$};

\draw[line width=1.5pt, color=black] (4,0.15) -- (4,-0.15);

\end{tikzpicture}
\end{center}
\caption{Example of an optimal allocation function under negative correlation when $\E[\omega]>\a$.}\label{fig:optimal_negative_allocation}
\end{figure}

\item {\em $\theta_H(\mu^*) = \BAR\theta$: almost all consumers do not consume in the private market.}

By \Cref{thm:optimal_negative}, consumers with types higher than $\theta_H(\mu^*)$ consume in the private market.  For these consumers, not only does the optimal allocation function coincide with the laissez-faire allocation, but the utility that they obtain must also coincide with their laissez-faire utility by the envelope theorem.  
However, since $\theta_H(\mu^*)=\BAR\theta$ when $\E[\omega]>\a$, the optimal redistribution program does not send consumers to the private market.
%
%
%

\item {\em $U^*(\theta)>U^{\LF}(\theta)$ for all $\theta<\BAR\theta$: outside of the free public option, consumers benefit from a nonlinear subsidy.}

In the proof of \Cref{thm:optimal_negative}, we show that only the \eqref{eq:IR} constraint of the highest consumer type can potentially bind; hence, the optimal redistribution program strictly benefits all consumers with lower types.  Consumers who do not consume the free public option purchase goods of higher quality, but these are nonetheless subsidized relative to the private market.
\end{enumerate}


Next, we explain \Cref{thm:optimal_negative} for the case when the average welfare weight of consumers does not exceed the welfare weight of profit, $\E[\omega]\leq\a$.  In this case, $\mu^*$ takes values between $\(\E[\omega]-\a\)_+=0$ and $\mu_{\max}$ (a similar argument to the one above ensures that $\mu^*$ is well-defined). 

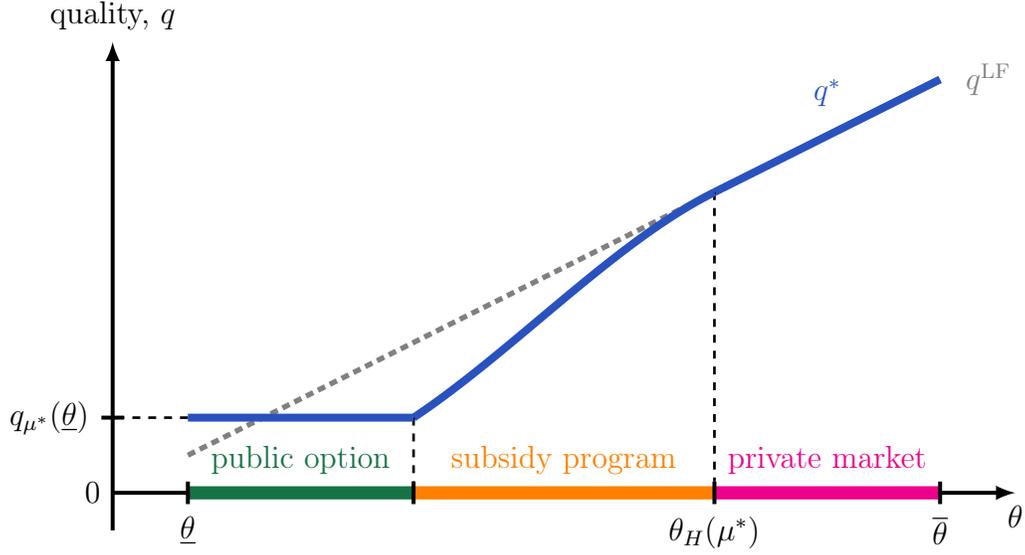
\begin{figure}
\begin{center}
\begin{tikzpicture}[scale=1.]

\draw[scale=1, domain=1:11., densely dashed, variable=\x, line width=2.pt, color=black!50] plot ({\x}, {0.5*\x});
\draw (11.2,5.5) node[right] {$\color{black!50}q^{\LF}$};

\draw[scale=1, domain=8:11., smooth, variable=\x, line width=3pt, color=ceruleanblue] plot ({\x}, {0.5*\x});
\draw[line width=1pt, dashed, color=black] (8,4) -- (8,0.15);
\draw (9.5,5) node[above] {$\color{ceruleanblue}q^*$};

\draw[scale=1, domain=4:8., smooth, variable=\x, line width=3pt, color=ceruleanblue] plot ({\x}, {-1*\x^3/48 + 17*\x^2/48 - 7*\x/6 + 4/3});
\draw[line width=1pt, dashed, color=black] (4,1) -- (4,0.15);

\draw[scale=1, domain=1:4., smooth, variable=\x, line width=3pt, color=ceruleanblue] plot ({\x}, {1});
\draw[line width=1pt, dashed, color=black] (1,1) -- (0.15,1);
\draw[line width=1.5pt, color=black] (0.15,1) -- (-0.15,1) node[left] {$q_{\mu^*}(\und\theta)$};

\draw (9.5,0.1) node [above] {{\color{magenta} private market}};

\draw (2.5,0.1) node [above] {{\color{darkspringgreen} public option}};
\draw (6,0.1) node [above] {{\color{orange} subsidy program}};

\draw [-{latex[scale=1.2]}, line width=1.5pt] (0,0) node [left] {0} -- (0,0) -- (12,0) node [below] {$\theta$};
\draw [-{latex[scale=1.2]}, line width=1.5pt] (0, -.5) -- (0,0) -- (0,6) node [above] {quality, $q$};

\draw[line width=5pt,magenta] (8,0) -- (11,0);
\draw[line width=5pt,darkspringgreen] (1,0) -- (4,0);
\draw[line width=5pt,orange] (4,0) -- (8,0);

\draw[line width=1.5pt,] (1,0.15) -- (1,-0.15) node[below] {$\und\theta$};
\draw[line width=1.5pt,] (11.,0.15) -- (11.,-0.15) node[below] {$\BAR\theta$};
\draw[line width=1.5pt, color=black] (8,0.15) -- (8,-0.15) node[below] {$\theta_H(\mu^*)$};
\draw[line width=1.5pt, color=black] (4,0.15) -- (4,-0.15);

\end{tikzpicture}
\end{center}
\caption{Example of an optimal allocation function under negative correlation when $\E[\omega]\leq\a$.}\label{fig:optimal_negative_allocation_another}

\end{figure}

We compare the optimal mechanism to the previous case by interpreting four properties when $\E[\omega]\leq\a$.  \Cref{fig:optimal_negative_allocation_another} shows an example of an optimal allocation function in this case.

\begin{enumerate}[label={\em(\roman*)}]
\item {\em The optimal redistribution program includes a free public option if and only if $\mu^*>0$.}

Like the previous case, $\mu^*>0$ implies that the optimal allocation function must be flat in a neighborhood of $\und\theta$ and that the \eqref{eq:LS} constraint binds; hence, we can similarly interpret this as a free public option.  However, unlike the previous case, it is possible that $\mu^*=0$, in which case no free public option is provided.

\item {\em Consumers with types between $\theta_H(\mu^*)$ and $\BAR\theta$ consume in the private market.}

Unlike the previous case, $\theta_H(\mu^*) < \BAR\theta$ in general when $\E[\omega]\leq\a$; and by \Cref{thm:optimal_negative}, consumers with types higher than $\theta_H(\mu^*)$ consume in the private market.  
%
%
%

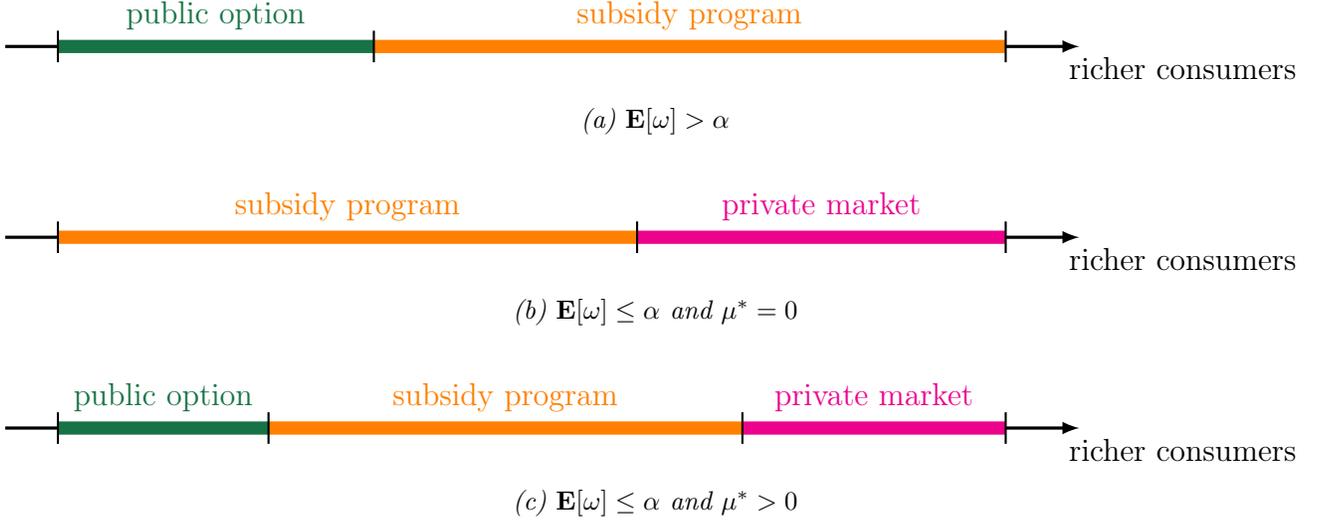
\begin{figure}[t!]
\centering
\begin{subfigure}[t]{\textwidth}
\centering
\begin{tikzpicture}[scale=1.4]
\draw[-latex,very thick] (-5,0) -- (5.2,0) ; 
\node[below right] at (5, 0.) {richer consumers};

\draw[line width=5pt,darkspringgreen] (-4.5,0) -- (-3, 0) node[above] {public option} -- (-1.5,0);
\draw[line width=5pt,orange] (-1.5,0) -- (1.5,0) node[above] {subsidy program} -- (4.5,0);
\draw[thick] (-4.5, 0.15) -- (-4.5, -0.15);
\draw[thick] (-1.5, 0.15) -- (-1.5, -0.15);
\draw[thick] (4.5, 0.15) -- (4.5, -0.15);

\end{tikzpicture}
\caption{$\E[\omega]>\a$}
\label{fig:negative_strong}
\end{subfigure}
\vspace{15pt}

\begin{subfigure}[t]{\textwidth}
\centering
\begin{tikzpicture}[scale=1.4]
\draw[-latex,very thick] (-5,0) -- (5.2,0) ; 
\node[below right] at (5, 0.) {richer consumers};

\draw[line width=5pt,orange] (-4.5,0) -- (-1.75,0) node[above] {subsidy program} -- (1,0);
\draw[line width=5pt,magenta] (1,0) -- (2.75,0) node[above] {private market} -- (4.5,0);
\draw[thick] (-4.5, 0.15) -- (-4.5, -0.15);
\draw[thick] (1, 0.15) -- (1, -0.15);
\draw[thick] (4.5, 0.15) -- (4.5, -0.15);

\end{tikzpicture}
\caption{$\E[\omega]\leq\a$ and $\mu^*=0$}
\label{fig:negative_weak_mu0}
\end{subfigure}
\vspace{15pt}

\begin{subfigure}[t]{\textwidth}
\centering
\begin{tikzpicture}[scale=1.4]
\draw[-latex,very thick] (-5,0) -- (5.2,0) ; 
\node[below right] at (5, 0.) {richer consumers};

\draw[line width=5pt,darkspringgreen] (-4.5,0) -- (-3.5,0) node[above] {public option} -- (-2.5,0);
\draw[line width=5pt,orange] (-2.5,0) -- (-0.25,0) node[above] {subsidy program} -- (2,0);
\draw[line width=5pt,magenta] (2,0) -- (3.25,0) node[above] {private market} -- (4.5,0);
\draw[thick] (-4.5, 0.15) -- (-4.5, -0.15);
\draw[thick] (-2.5, 0.15) -- (-2.5, -0.15);
\draw[thick] (2, 0.15) -- (2, -0.15);
\draw[thick] (4.5, 0.15) -- (4.5, -0.15);

\end{tikzpicture}
\caption{$\E[\omega]\leq\a$ and $\mu^*>0$}
\label{fig:negative_weak}
\end{subfigure}
\vspace{5pt}

\caption{Optimal in-kind redistribution programs under negative correlation.}\label{fig:negative}
\end{figure}

\item {\em Consumers in the middle benefit from a nonlinear subsidy.}

Similar to the previous case, consumers who do not consume in the private market can still purchase goods of higher quality than the public option, which are subsidized relative to the private market.

\item {\em For consumers with types below $\theta_H(\mu^*)$, quality consumption is distorted upwards for poorer consumers and downwards for richer consumers.}

The sign of the distortion for a consumer of type $\theta\leq\theta_H(\mu^*)$ is the same as the sign of the difference $H_{\mu^*}(\theta)-\theta$.  In turn, the latter shares the same sign as
\begin{align*}
\left[H_{\mu^*}(\theta)-\theta\right]\cdot\a f(\theta)
&=\mu^* \und\theta\cdot\d_{\und\theta}(\theta) + \mu^* + \int_{\und\theta}^\theta\left[\a-\omega(s)\right]\ \dd F(s).
\end{align*}
Given that $\omega$ is decreasing, it follows that $\theta\mapsto\int_{\und\theta}^\theta\left[\a-\omega(s)\right]\ \dd F(s)$ is quasiconvex; hence the above expression is quasiconvex in $\theta$
.  Moreover, since $H_{\mu^*}(\theta_H(\mu^*))=\theta_H(\mu^*)$, it follows that there exists $\hat\theta\in[\und\theta,\theta_H(\mu^*)]$ such that $q^*(\theta) > q^{\LF}(\theta)$ for $\und\theta\leq\theta <\hat\theta$ and $q^*(\theta) < q^{\LF}(\theta)$ for $\hat\theta<\theta \leq\BAR\theta$.
\end{enumerate}

We summarize the possible structures of the optimal in-kind redistribution program under negative correlation in \Cref{fig:negative_strong,fig:negative_weak_mu0,fig:negative_weak}.

\subsection{Optimal Mechanisms Under Positive Correlation}

Next, we characterize the optimal mechanism when welfare weight is positively correlated with willingness to pay.

\begin{theorem}[characterization of optimal mechanisms under positive correlation]\label{thm:optimal_positive}
Suppose that $\omega$ is increasing.  For any $\mu\geq0$, define 
\[q_\mu(\theta) \coloneq D\(c,\BAR{J_\mu}(\theta)\),\quad\text{where }J_\mu(\theta)\coloneq \theta + \frac{\mu\und\theta\cdot \d_{\und\theta}(\theta) - \int_{\theta}^{\BAR\theta}\left[\a-\omega(s)\right]\ \dd F(s)}{\a f(\theta)}.\]
Moreover, denote
\[\theta_L^*\coloneq\min\left\{\theta\in[\und\theta,\BAR\theta]:\int_{\theta}^{\BAR\theta}\left[\a-\omega(s)\right]\ \dd F(s)\leq0\right\}.\]
Let $\mu^* \coloneq \(\E[\omega]-\a\)_+$.  Then the optimal allocation function is 
\[q^*(\theta) = \begin{dcases}
q_{\mu^*}(\theta) &\text{for }\theta_L^*\leq \theta\leq\BAR\theta,\\
q^{\LF}(\theta) &\text{for }\und\theta\leq \theta < \theta_L^*.
\end{dcases}\]
\end{theorem}

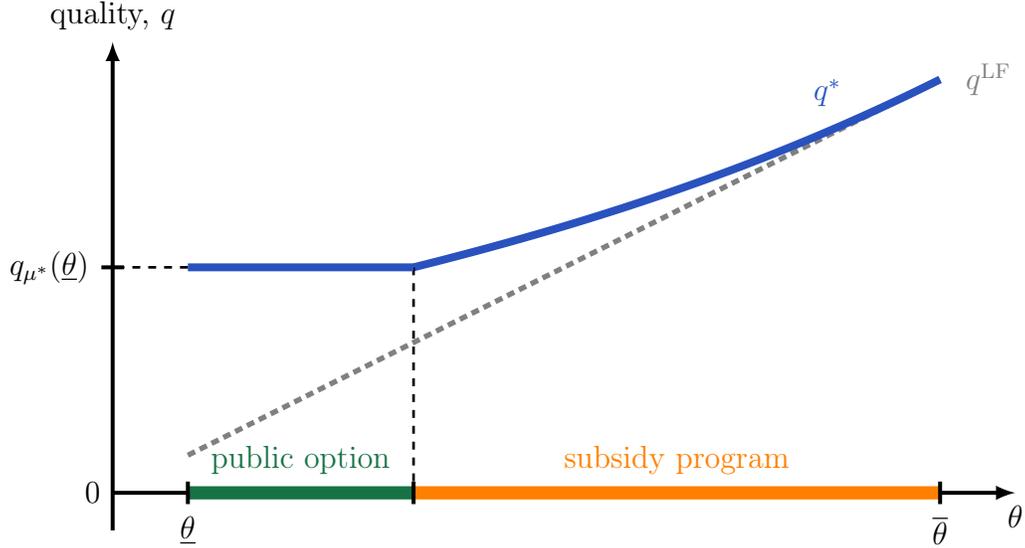
\begin{figure}
\begin{center}
\begin{tikzpicture}[scale=1.]

\draw[scale=1, domain=1:11., densely dashed, variable=\x, line width=2.pt, color=black!50] plot ({\x}, {0.5*\x});
\draw (11.2,5.5) node[right] {$\color{black!50}q^{\LF}$};

\draw (9.5,5) node[above] {$\color{ceruleanblue}q^*$};

\draw[scale=1, domain=4:11., smooth, variable=\x, line width=3pt, color=ceruleanblue] plot ({\x}, {\x^3/1372+\x^2/686+279*\x/1372+726/343});
\draw[line width=1pt, dashed, color=black] (4,3) -- (4,0.15);

\draw[scale=1, domain=1:4., smooth, variable=\x, line width=3pt, color=ceruleanblue] plot ({\x}, {3});
\draw[line width=1pt, dashed, color=black] (1,3) -- (0.15,3);
\draw[line width=1.5pt, color=black] (0.15,3) -- (-0.15,3) node[left] {$q_{\mu^*}(\und\theta)$};

\draw (2.5,0.1) node [above] {{\color{darkspringgreen} public option}};
\draw (7.5,0.1) node [above] {{\color{orange} subsidy program}};

\draw [-{latex[scale=1.2]}, line width=1.5pt] (0,0) node [left] {0} -- (0,0) -- (12,0) node [below] {$\theta$};
\draw [-{latex[scale=1.2]}, line width=1.5pt] (0, -.5) -- (0,0) -- (0,6) node [above] {quality, $q$};

\draw[line width=5pt,darkspringgreen] (1,0) -- (4,0);
\draw[line width=5pt,orange] (4,0) -- (11,0);

\draw[line width=1.5pt,] (1,0.15) -- (1,-0.15) node[below] {$\und\theta$};
\draw[line width=1.5pt,] (11.,0.15) -- (11.,-0.15) node[below] {$\BAR\theta$};

\draw[line width=1.5pt, color=black] (4,0.15) -- (4,-0.15) node[below] {};

\end{tikzpicture}
\end{center}
\caption{Example of an optimal allocation function under positive correlation when $\E[\omega]>\a$.}\label{fig:optimal_positive_allocation}

\end{figure}

To explain \Cref{thm:optimal_positive}, we highlight three properties of the optimal mechanism in the case where the average welfare weight of consumers exceeds the welfare weight of profit, $\E[\omega]>\a$.  \Cref{fig:optimal_positive_allocation} shows an example of the optimal allocation function in this case.

\begin{enumerate}[label={\em(\roman*)}]
\item {\em $\mu^*>0$: the optimal redistribution program always includes a free public option.}

Observe that $\mu^*=\(\E[\omega]-\a\)_+>0$ when $\E[\omega]>\a$.  This means that $J_{\mu^*}$ has an atom at $\und\theta$; hence $\BAR{J_{\mu^*}}$---and, by extension, $q^*=q_{\mu^*}$---must be flat in a neighborhood of $\und\theta$.  Our proof of \Cref{thm:optimal_positive} also shows that $\mu^*$ is the shadow cost of the \eqref{eq:LS} constraint.  As in the case of \Cref{thm:optimal_negative}, we interpret the resulting flat region of $q^*$ as a free public option.

\item {\em $\theta_L^* = \und\theta$: almost all consumers do not consume in the private market.}

By \Cref{thm:optimal_positive}, consumers with types lower than $\theta_L^*$ consume in the private market.  However, observe that
\[\int_{\und\theta}^{\BAR\theta}\left[\a-\omega(s)\right]\ \dd F(s) < 0\implies \theta_L^* = \und\theta.\]
Consequently, the optimal redistribution program does not send consumers to the private market when $\E[\omega]>\a$.

\item {\em No participation constraints bind: the social planner is not constrained by consumers' ability to access the private market.}

In the proof of \Cref{thm:optimal_positive}, we show that the social planner is not additionally constrained by the ability of consumers to access the private market in this case: \eqref{eq:IR} constraints do not bind.  Consumers who do not consume the public option benefit from a nonlinear subsidy.
\end{enumerate}

Next, we explain \Cref{thm:optimal_positive} for the case when the average welfare weight of consumers does not exceed the welfare weight of profit, $\E[\omega]\leq\a$.  We highlight four properties in this case.  \Cref{fig:optimal_positive_allocation_another} shows an example of an optimal allocation function.

\begin{enumerate}[label={\em(\roman*)}]
\item {\em The optimal redistribution program never includes a free public option.}

In this case, $\mu^*=\(\E[\omega]-\a\)_+ =0$ as $\E[\omega]\leq\a$.  This means that a free public option is never provided.  Intuitively, this is because a free public option benefits consumers with the lowest consumer types, who are precisely the richest consumers in this case.

\item {\em Consumers with types between $\und\theta$ and $\theta_L^*$ consume in the private market.}

Unlike the previous case, $\theta_L^*> \und\theta$ when $\E[\omega]<\a$.  This is because $\theta\mapsto \int_\theta^{\BAR\theta}\left[\a-\omega(s)\right]\ \dd F(s)$ is quasiconvex and equal to zero at $\BAR\theta$; however, when $\E[\omega]<\a$, it is positive at $\und\theta$.  Thus $\theta_L^*>\und\theta$; consumers with types below $\theta_L^*$ consume in the private market.

%
%
%

\item {\em Consumers who do not consume in the private market benefit from a nonlinear subsidy.}

Similar to the previous case, consumers who do not consume in the private market can still purchase goods of higher quality, which are subsidized relative to the private market.

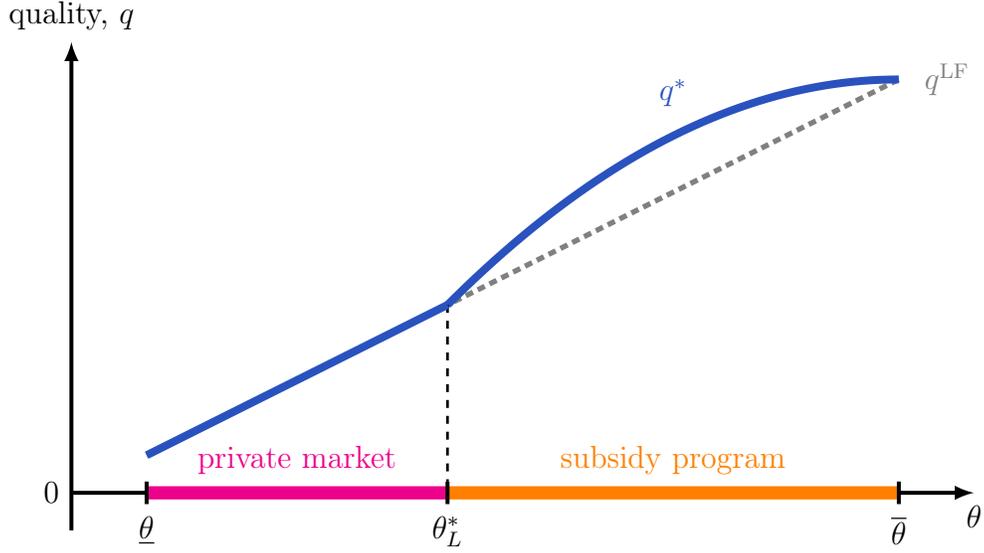
\begin{figure}[t!]
\begin{center}
\begin{tikzpicture}[scale=1]

\draw[scale=1, domain=1:11., densely dashed, variable=\x, line width=2.pt, color=black!50] plot ({\x}, {0.5*\x});
\draw (11.2,5.5) node[right] {$\color{black!50}q^{\LF}$};

\draw[scale=1, domain=5:11., smooth, variable=\x, line width=3pt, color=ceruleanblue] plot ({\x}, {-\x^2/12+11*\x/6-55/12});
\draw (8,5) node[above] {$\color{ceruleanblue}q^*$};

\draw[line width=1pt, dashed, color=black] (5,2.5) -- (5,0.15);

\draw[scale=1, domain=1:5., smooth, variable=\x, line width=3pt, color=ceruleanblue] plot ({\x}, {0.5*\x});


\draw (3,0.1) node [above] {{\color{magenta} private market}};
\draw (8,0.1) node [above] {{\color{orange} subsidy program}};

\draw [-{latex[scale=1.2]}, line width=1.5pt] (0,0) node [left] {0} -- (0,0) -- (12,0) node [below] {$\theta$};
\draw [-{latex[scale=1.2]}, line width=1.5pt] (0, -.5) -- (0,0) -- (0,6) node [above] {quality, $q$};

\draw[line width=5pt,magenta] (1,0) -- (5,0);
\draw[line width=5pt,orange] (5,0) -- (11,0);

\draw[line width=1.5pt,] (1,0.15) -- (1,-0.15) node[below] {$\und\theta$};
\draw[line width=1.5pt,] (11.,0.15) -- (11.,-0.15) node[below] {$\BAR\theta$};
\draw[line width=1.5pt, color=black] (5,0.15) -- (5,-0.15) node[below] {$\theta_L^*$};

\end{tikzpicture}
\end{center}
\caption{Example of an optimal allocation function under positive correlation when $\E[\omega]\leq\a$.}\label{fig:optimal_positive_allocation_another}

\end{figure}

\item {\em For consumers with types between $\theta_L^*$ and $\BAR\theta$, quality consumption is distorted upwards.}

The sign of the distortion for a consumer of type $\theta\geq\theta_L^*$ is the same as the sign of the difference $J_{\mu^*}(\theta)-\theta$.  In turn, the latter shares the same sign as
\begin{align*}
\left[J_{\mu^*}(\theta)-\theta\right]\cdot\a f(\theta)
&=\mu^* \und\theta\cdot\d_{\und\theta}(\theta) - \int_{\theta}^{\BAR\theta}\left[\a-\omega(s)\right]\ \dd F(s).
\end{align*}
Given that $\omega$ is increasing, it follows that $\theta\mapsto\int_{\theta}^{\BAR\theta}\left[\a-\omega(s)\right]\ \dd F(s)$ is quasiconvex; hence the above expression is quasiconcave in $\theta$
.  Moreover, since $J_{\mu^*}(\theta_L^*)=\theta_L^*$ and $J_{\mu^*}(\BAR\theta)=\BAR\theta$, it follows that the consumption of consumers with types between $\theta_L^*$ and $\BAR\theta$ is distorted upwards.
\end{enumerate}

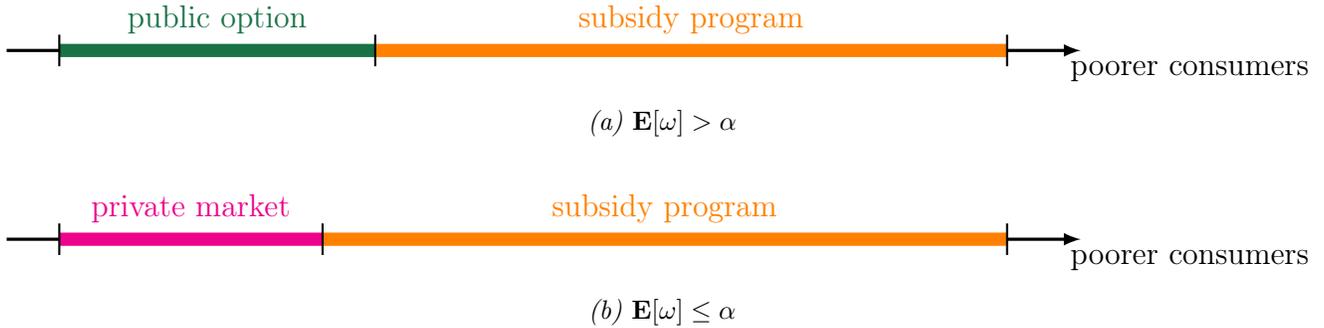
\begin{figure}[t!]
\centering
\begin{subfigure}[t]{\textwidth}
\centering
\begin{tikzpicture}[scale=1.4]
\draw[-latex,very thick] (-5,0) -- (5.2,0) ; 
\node[below right] at (5, 0.) {poorer consumers};

\draw[line width=5pt,darkspringgreen] (-4.5,0) -- (-3, 0) node[above] {public option} -- (-1.5,0);
\draw[line width=5pt,orange] (-1.5,0) -- (1.5,0) node[above] {subsidy program} -- (4.5,0);
\draw[thick] (-4.5, 0.15) -- (-4.5, -0.15);
\draw[thick] (-1.5, 0.15) -- (-1.5, -0.15);
\draw[thick] (4.5, 0.15) -- (4.5, -0.15);

\end{tikzpicture}
\caption{$\E[\omega]>\a$}
\label{fig:positive_strong}
\end{subfigure}
\vspace{15pt}

\begin{subfigure}[t]{\textwidth}
\centering
\begin{tikzpicture}[scale=1.4]
\draw[-latex,very thick] (-5,0) -- (5.2,0) ; 
\node[below right] at (5, 0.) {poorer consumers};

\draw[line width=5pt,magenta] (-4.5,0) -- (-3.25,0) node[above] {private market} -- (-2.,0);
\draw[line width=5pt,orange] (-2.,0) -- (1.25,0) node[above] {subsidy program} -- (4.5,0);
\draw[thick] (-4.5, 0.15) -- (-4.5, -0.15);
\draw[thick] (-2., 0.15) -- (-2., -0.15);
\draw[thick] (4.5, 0.15) -- (4.5, -0.15);

\end{tikzpicture}
\caption{$\E[\omega]\leq\a$}
\label{fig:positive_weak}
\end{subfigure}
\vspace{5pt}

\caption{Optimal in-kind redistribution programs under positive correlation.}\label{fig:positive}
\end{figure}

We summarize the possible structures of the optimal in-kind redistribution program under positive correlation in \Cref{fig:positive_strong,fig:positive_weak}.

\subsection{Intuition for Characterization of Optimal Mechanisms}

In \ref{app:proof_main}, we prove \Cref{thm:optimal_negative,thm:optimal_positive} using a Lagrangian approach {\em \`a la} \cite{amadorbagwell13}.  This is done in three steps: {\em(i)} guessing the binding participation constraints (\ie, guessing their Lagrange multipliers); {\em(ii)} solving the relaxed problem by maximizing the Lagrangian; and {\em(iii)} verifying optimality in the original problem using the \cite{luenberger69} sufficiency theorem.

Here, we provide intuition for our proofs of \Cref{thm:optimal_negative,thm:optimal_positive}.  To this end, we introduce results below that help shed light on the structure of optimal mechanisms.  While none of these results are required for the proofs of \Cref{thm:optimal_negative,thm:optimal_positive}, they help to motivate our guesses of the binding participation constraints in \ref{app:proof_main}.

\subsubsection{Intuition for Optimal Mechanisms Under Negative Correlation}

We first consider which participation constraints bind when welfare weight is negatively correlated with willingness to pay.

First, it is possible that no participation constraints bind.  In this case, we can solve the social planner's problem as we would a ``full mechanism design'' problem.  As we show in \ref{app:proof_main}, this turns out to be the case sometimes when $\E[\omega]>\a$.

Second, it is possible that participation constraints bind for an interval of types, say $(\theta_-,\theta_+)$.  In this case, it is straightforward to show that each of these types is allocated his laissez-faire allocation: $q^*(\theta)=q^{\LF}(\theta)$ for any $\theta\in(\theta_-,\theta_+)$.  We interpret consumers in this interval as consuming in the private market, and we say that the private market is ``active.'' 

\begin{lemma}\label{lem:active_negative_1}
Whenever the private market is active, the \eqref{eq:IR} constraint binds for the consumer with the highest type, $\BAR\theta$.
\end{lemma}

\Cref{lem:active_negative_1} is intuitive: the social planner has no reason to subsidize the richest consumers if she does not already subsidize a positive measure of relatively poorer consumers.  

\begin{lemma}\label{lem:active_negative_2}
There are no ``gaps'' between intervals on which the \eqref{eq:IR} constraint binds: if the \eqref{eq:IR} constraint binds on the intervals $(\theta_-,\theta_+)$ and $(\theta_-',\theta_+')$, then the \eqref{eq:IR} constraint in fact binds on the interval $(\min\{\theta_-,\theta_-'\},\max\{\theta_+,\theta_+'\})$.
\end{lemma}

\Cref{lem:active_negative_2} extends the logic of \Cref{lem:active_negative_1}: the social planner has no reason to subsidize consumers in the ``gaps'' between those who consume in the private market. 

Together, \Cref{lem:active_negative_1,lem:active_negative_2} imply that, whenever the private market is active, an interval of consumers that includes the highest type $\BAR\theta$ purchases goods from the private market.  While this does not rule out the \eqref{eq:IR} constraint binding for individual consumers outside of this interval, we guess---and verify---that this can never be the case in our proof in \ref{app:proof_main}.  We formally prove \Cref{lem:active_negative_1,lem:active_negative_2} in \ref{app:additional_proofs} using necessary conditions in the Lagrangian approach.

Finally, it is possible that participation constraints bind for some consumers even when the private market is inactive.  Motivated by \Cref{lem:active_negative_1}, we guess that, in this case, the \eqref{eq:IR} constraint can only bind for the consumer with the highest type, $\BAR\theta$.  In \ref{app:proof_main}, we show that this is the case sometimes when $\E[\omega]>\a$.

\subsubsection{Intuition for Optimal Mechanisms Under Positive Correlation}

Next, we consider which participation constraints bind when welfare weight is positively correlated with willingness to pay.

Again, it is possible that no participation constraints bind.  As we show in \ref{app:proof_main}, this turns out to be the case if and only if $\E[\omega]>\a$.

When the private market is active, we can show that analogs of \Cref{lem:active_negative_1} and \Cref{lem:active_negative_2} hold.  Specifically, the \eqref{eq:IR} constraint binds for the consumer with the lowest type, $\und\theta$, since that consumer is the richest consumer when welfare weight is positively correlated with willingness to pay.  In addition, \Cref{lem:active_negative_2} holds in this setting; hence, whenever the private market is active, an interval of consumers that includes the lowest type $\und\theta$ consumes in the private market.

Finally, we show in \ref{app:proof_main} that participation constraints cannot bind for any consumer when the private market is inactive.

\section{Economic Implications}
\label{sec:implications}

We now use our main results to shed light on four aspects of optimal in-kind redistribution: {\em(i)}~identifying which consumers benefit from optimal intervention; {\em(ii)} determining when non-market allocations in the form of a free public option are justified; {\em(iii)} understanding how the social planner's opportunity cost of budget affects optimal redistribution; and {\em(iv)} establishing when consumers should be prevented from topping up their consumption in the private market.  We defer all proofs to \ref{app:additional_proofs}.

\subsection{Who Benefits From Optimal Intervention?}

We begin by using \Cref{thm:optimal_negative,thm:optimal_positive} to characterize which consumers strictly benefit from the optimal mechanism.

\subsubsection{Negative Correlation}

We first study the case in which consumers' welfare weights are negatively correlated with their willingness to pay.

\begin{proposition}[beneficiaries of optimal intervention under negative correlation]\label{prop:benefit_negative}
Suppose that $\omega$ is decreasing.  Let $\theta_H$, $\mu^*$, and $q_\mu$ be as defined in \Cref{thm:optimal_negative}.
\begin{enumerate}[label={(\roman*)}]
\item If $\E[\omega]\leq\a$, then consumers with types $\theta<\theta_H(\mu^*)$ benefit from the optimal mechanism.
\item If $\E[\omega]>\a$, then all consumers benefit from the optimal mechanism if
\[\und\theta v(q_{\E[\omega]-\a}(\und\theta)) + \int_{\und\theta}^{\BAR\theta} v(q_{\E[\omega]-\a}(s))\ \dd s > \und U^{\LF} + \int_{\und\theta}^{\BAR\theta} v(q^{\LF}(s))\ \dd s.\]
Otherwise, all consumers except the highest type $\BAR\theta$ benefit from the optimal mechanism.
\end{enumerate}
\end{proposition}

While \Cref{prop:benefit_negative} follows straightforwardly from \Cref{thm:optimal_negative}, we compare our result to a counterfactual setting where the social planner controls the entire market.  Specifically, suppose that the \eqref{eq:IR} constraints are replaced by the restriction that $\und U\geq0$.  As we show in \ref{app:proof_main}:
\begin{enumerate}[label={\em(\roman*)}]
\item If $\E[\omega]\leq\a$, then all consumers are worse off under the counterfactual optimal mechanism relative to the laissez-faire outcome.

\item If $\E[\omega]>\a$, then, as in \Cref{prop:benefit_negative}, all consumers benefit from the counterfactual optimal mechanism if
\[\und\theta v(q_{\E[\omega]-\a}(\und\theta)) + \int_{\und\theta}^{\BAR\theta} v(q_{\E[\omega]-\a}(s))\ \dd s > \und U^{\LF} + \int_{\und\theta}^{\BAR\theta} v(q^{\LF}(s))\ \dd s.\]
Otherwise, unlike \Cref{prop:benefit_negative}, let $\hat\theta$ be the consumer type defined by 
\[\hat\theta \coloneq \min\left\{\theta\in[\und\theta,\BAR\theta]:\und\theta v(q_{\E[\omega]-\a}(\und\theta)) + \int_{\und\theta}^{\theta} v(q_{\E[\omega]-\a}(s))\ \dd s \leq \und U^{\LF} + \int_{\und\theta}^{\theta} v(q^{\LF}(s))\ \dd s\right\}.\]
Then, consumers with types $\theta<\hat\theta$ benefit from the counterfactual optimal mechanism, while consumers with types $\theta>\hat\theta$ are worse off.
\end{enumerate}

This comparison shows how the social planner's response to binding participation constraints varies depending on whether the average consumer welfare weight is higher than the welfare weight of profit.  If $\E[\omega]\leq\a$, despite the fact that {\em all} participation constraints are violated under the counterfactual optimal mechanism, the social planner responds by targeting the redistribution program to poorer consumers.  In this case, the social planner responds to binding participation constraints by making some consumers indifferent between the optimal mechanism and the laissez-faire outcome and restricting subsidies to the poorer consumers.  By contrast, if $\E[\omega]>\a$, the social planner does not respond by targeting the redistribution program to poorer consumers, and subsidies remain unrestricted.  Instead, the social planner makes all consumers (except the highest type $\BAR\theta$) better off because sufficiently large benefits of doing so accrue to poorer consumers. 

\subsubsection{Positive Correlation}

We now consider the case in which consumers' welfare weights are positively correlated with their willingness to pay.
 
\begin{proposition}[beneficiaries of optimal intervention under positive correlation]\label{prop:benefit_positive}
Suppose that $\omega$ is increasing.  Let $\theta_L^*$ be as defined in \Cref{thm:optimal_positive}.
\begin{enumerate}[label={(\roman*)}]
\item If $\E[\omega]\leq\a$, then consumers with types $\theta>\theta_L^*$ benefit from the optimal mechanism.
\item If $\E[\omega]>\a$, then all consumers benefit from the optimal mechanism.
\end{enumerate}
\end{proposition}

While \Cref{prop:benefit_positive} follows straightforwardly from \Cref{thm:optimal_positive}, we again compare our result to the same counterfactual setting where the social planner controls the entire market.  As we show in \ref{app:proof_main}:
\begin{enumerate}[label={\em(\roman*)}]
\item If $\E[\omega]\leq\a$, then all consumers are worse off under the counterfactual optimal mechanism relative to the laissez-faire outcome if
\[\int_{\und\theta}^{\BAR\theta}v(q_{\E[\omega]-\a}(s))\ \dd s < \und U^{\LF} + \int_{\und\theta}^{\BAR\theta} v(q^{\LF}(s))\ \dd s.\]
Otherwise, let $\hat\theta$ be the consumer type defined by
\[\hat\theta\coloneq \min\left\{\theta\in[\und\theta,\BAR\theta]:\int_{\und\theta}^{\theta}v(q_{\E[\omega]-\a}(s))\ \dd s \geq \und U^{\LF} + \int_{\und\theta}^{\theta}v(q^{\LF}(s))\ \dd s\right\}.\]
Then consumers with types $\theta<\hat\theta$ are worse off under the counterfactual optimal mechanism, while consumers with types $\theta>\hat\theta$ benefit.

\item If $\E[\omega]>\a$, then, as in \Cref{prop:benefit_positive}, all consumers benefit from the counterfactual optimal mechanism.\end{enumerate}

Because participation constraints can bind only if $\E[\omega]\leq\a$, this comparison shows that the social planner's response is to target the redistribution program to poorer consumers in this case.  While participation constraints are violated for types $\theta<\hat\theta$ under the counterfactual optimal mechanism, it can be shown that $\theta_L^*<\hat\theta$.  Consequently, not all consumers whose participation constraints are violated under the counterfactual optimal mechanism are made indifferent by the social planner: a positive measure of consumers with types between $\theta_L^*$ and $\hat\theta$ are made strictly better off.

Together, \Cref{prop:benefit_negative,prop:benefit_positive} highlight that when consumers can access a private market, the social planner targets the optimal redistribution program by restricting it to poorer consumers---but only if $\E[\omega]\leq\a$.  This contrasts with \Cref{thm:scope}: even though the social planner optimally intervenes whenever $\max\omega >\a$, this intervention is limited to only some consumers in the market when $\E[\omega]\leq\a$.

\subsection{When Should Non-Market Allocations Be Used?}

Next, we use \Cref{thm:optimal_negative,thm:optimal_positive} to characterize when non-market allocations---in the form of a free public option---should be used.

\begin{proposition}[optimality of non-market allocations]\label{prop:non-market}
Let $\theta_H$ and $q_\mu$ be as defined in \Cref{thm:optimal_negative}.
\begin{enumerate}[label={(\roman*)}]
\item When $\omega$ is decreasing, a free public option is optimal if and only if either {\em(a)} $\E[\omega]>\a$, or {\em(b)} $\E[\omega]\leq\a$ and
\[\int_{\und\theta}^{\theta_H(0)}v(q_0(s))\ \dd s + \und\theta v(q_0(\und\theta)) < U^{\LF}(\theta_H(0)).\]

\item When $\omega$ is increasing, a free public option is optimal if and only if $\E[\omega]>\a$.
\end{enumerate}
\end{proposition}

\Cref{prop:non-market} shows that, when welfare weight is negatively correlated with willingness to pay, the ability of consumers to access a private market broadens the justification for using non-market allocations.  Indeed, as we show in \ref{app:proof_main}, a social planner who can control the entire market uses a free public option if and only if $\E[\omega]>\a$.  When consumers have access to a private market, a free public option remains optimal in this case, but it may also be optimal even when $\E[\omega]\leq\a$ under the additional condition specified in \Cref{prop:non-market}.  As the condition formalizes, this broadened justification for non-market allocations arises because a free public option may help to relax participation constraints that would otherwise be binding for poorer consumers.

\subsection{How Does Budget Affect Optimal Redistribution?}

We now turn our attention to how the social planner's outside options---reflected in her opportunity cost of budget, $\a$---affect the design of her optimal redistribution program.

A higher opportunity cost of budget clearly reduces the scope of redistribution.  On one hand, this is intuitive: as alternative redistribution programs available to the social planner become more effective, she increasingly prefers to rely on those options.  On the other hand, this follows directly from \Cref{thm:scope}: as $\a$ increases, the condition $\max\omega>\a$ becomes less likely to hold, reducing the likelihood that the social planner can strictly improve upon the laissez-faire outcome.

However, when the social planner still chooses to intervene, it is less clear how a higher opportunity cost of budget impacts the structure of the optimal redistribution program.  To address this, we analyze the comparative statics of the optimal mechanisms described by \Cref{thm:optimal_negative,thm:optimal_positive} with respect to $\a$.

\begin{proposition}[comparative statics of optimal mechanisms with respect to $\a$]\label{prop:comp_statics}
Let $\theta_H$, $\theta_L^*$, and $\mu^*$ be as defined in \Cref{thm:optimal_negative,thm:optimal_positive}.
\begin{enumerate}[label={(\roman*)}]
\item Suppose that $\omega$ is decreasing.  Then $\theta_H(\mu^*)$ increases with $\a$ and $\mu^*$ decreases with $\a$.\label{it:comp_statics_negative}
\item Suppose that $\omega$ is increasing.  Then both $\theta_L^*$ and $\mu^*$ decrease with $\a$.\label{it:comp_statics_positive}
\end{enumerate}
\end{proposition}

This result implies that as the opportunity cost of budget increases, the optimal redistribution program benefits fewer consumers (cf.~\Cref{prop:benefit_negative,prop:benefit_positive}) and reduces the scope of non-market allocations (cf.~\Cref{prop:non-market}).  The first result is intuitive: as the social planner's outside options improve, she reduces in-kind redistribution to {\em each} consumer, relying more on other programs captured by $\a$.  The second result, however, is less straightforward and demonstrates how a free public option acts as a substitute for the social planner's other available redistribution programs.


While our analysis has focused on comparative statics with respect to $\a$, similar results apply when considering $\omega$, which we interpret as the strength of the social planner's redistribution motive.  Specifically, a social planner who assigns a welfare weight function $\omega_H$ to consumers is said to have a stronger redistribution motive than one who assigns $\omega_L$, if $\omega_H(\theta)\geq \omega_L(\theta)$ for all $\theta\in[\und\theta,\BAR\theta]$.  Similar to the discussion above, a social planner with a stronger redistribution motive optimally intervenes more frequently, benefits more consumers when she does, and utilizes a free public option more often.

\subsection{When Should Topping Up Be Prevented?}

Finally, we examine the question of when consumers should be prevented from topping up their consumption in the private market.  

The issue of topping up has received some attention in the literature on in-kind redistribution.  While many studies focus exclusively on either the allowance or prohibition of topping up (\citealp{curriegahvari08}, provide a comprehensive survey), \cite{blomquistchristiansen98} offer a comparative analysis of both approaches.  Using a Mirrleesian model where consumers with different wage rates trade off labor and leisure, \citeauthor{blomquistchristiansen98} show that topping up should be allowed if and only if the demand for the publicly provided good decreases with leisure.

The issue of topping up is also relevant to public policy.  For instance, motivated at least in part by concerns about unequal access to education due to wealth and socioeconomic privilege, China launched strict regulations that made it  impractical for private tutors to operate \citep{palmer21}.  As such, consumers can no longer top up their consumption of education in the private tutoring market, and must instead choose between attending a public school or a private school.  South Korea has previously attempted to ban private tutoring due to concerns about educational inequality \citep{chandler11}, while Singapore is now considering related policies to address similar concerns \citep{zhangmokhtar24}.

In this subsection, we show that topping up should be prevented when welfare weight is negatively correlated with willingness to pay, but not when they are positively correlated.  Our mechanism-design approach complements the work of \citeauthor{blomquistchristiansen98} in two key ways.  First, we focus on an individual market where redistribution is driven by the correlation between consumers' heterogeneous consumption preferences and exogenously determined welfare weights, rather than by the interaction between consumption and labor market decisions.\footnote{As \cite{paistrack24} show, optimal mechanisms derived for consumption in individual markets can also be endogenized in a richer model with labor choice and income taxation.}  Second, we characterize optimal mechanisms in a richer environment, where the social planner can design a price schedule for different quality levels of the publicly provided good, rather than simply deciding whether or not to offer the good at a fixed level and price.

In our setting, the social planner never benefits from allowing consumers to top up.  This arises from the fact that the social planner faces tighter participation constraints when consumers are given the option to top up their public allocation by consuming in the private market.  As we show in our companion paper \citep{kangwatt24a}, the participation constraints when consumers are allowed to top up are given by $\und U\geq \und U^{\LF}$ and 
\begin{equation}\label{eq:IR'}
    q(\theta) \geq q^{\LF}(\theta)\qquad\text{for any }\theta\in[\und\theta,\BAR\theta].\tag{IR'}
\end{equation}
In particular, the total consumption $q(\theta)$ of each consumer must be at least equal to his laissez-faire consumption $q^{\LF}(\theta)$: whenever his public allocation is less than his laissez-faire consumption, he chooses to top up his consumption in the private market.  Because $\und U\geq \und U^{\LF}$, the \eqref{eq:IR'} constraints above imply the \eqref{eq:IR} constraints by the envelope theorem (cf.~\citealp{milgromsegal02}).  In turn, the tighter \eqref{eq:IR'} constraints further restrict the social planner's ability to redistribute.

Given that the social planner always weakly benefits from preventing topping up, we examine when the social planner strictly benefits from doing so:

\begin{proposition}[optimal prevention of topping up]\label{prop:topping}\hfill
\begin{enumerate}[label={(\roman*)}]
\item When $\omega$ is decreasing, the social planner strictly benefits from preventing topping up if and only if $\max\omega>\a$.\label{it:topping_negative}
\item When $\omega$ is increasing, the social planner never benefits from preventing topping up.\label{it:topping_positive}
\end{enumerate}
\end{proposition}

\Cref{prop:topping} states that when welfare weight is negatively correlated with willingness to pay, the social planner strictly benefits from preventing topping up whenever there is scope for in-kind redistribution; however, when welfare weight is positively correlated with willingness to pay, the social planner derives no benefit from preventing topping up.  Part~\ref{it:topping_negative} of \Cref{prop:topping} follows from \Cref{thm:scope,thm:optimal_negative}: the optimal allocation characterized in \Cref{thm:optimal_negative} is always distorted downwards for types just below $\theta_H(\mu^*)$, so the social planner must strictly benefit from preventing topping up whenever there is scope for in-kind redistribution, as characterized by \Cref{thm:scope}.  Part~\ref{it:topping_positive} of \Cref{prop:topping} then follows from the fact that the optimal allocation characterized in \Cref{thm:optimal_positive} satisfies the tighter \eqref{eq:IR'} constraints.

Consequently, \Cref{prop:topping} suggests that a positive correlation between welfare weight and willingness to pay not only ensures that in-kind redistribution programs are self-targeting, but also reduces the need for monitoring to prevent consumers from topping up.  Thus, this may help shed light on why many governments, especially in developing countries (which may lack the monitoring technology required to prevent topping up), choose to redistribute by subsidizing inferior goods like public transit, agricultural products, and basic staples such as coarse bread and cassava.\footnote{Subsidies for inferior goods have also long been advocated by both economists and policymakers.  For example, in the {\em Public Expenditure Handbook: A Guide to Public Policy Issues in Developing Countries} published by the International Monetary Fund, \cite{mackenzie91} recommends that ``[m]arketed goods with a negative income elasticity (i.e., inferior goods) are ideal candidates for a redistributive subsidy.''}  While the self-targeting property of subsidizing inferior goods has long been recognized (see, \eg, \citealp{nicholszeckhauser82}), our \Cref{prop:topping} proposes a complementary explanation: when welfare weight is positively correlated with willingness to pay (as is often the case for inferior goods), governments can maximize the effectiveness of in-kind redistribution without needing to allocate resources to prevent consumers from topping up in private markets. 

\section{Discussion}
\label{sec:discussion}

To illustrate the practical implications and limitations of our findings, we now apply our results to real-world applications of in-kind redistribution.  Specifically, we focus on: {\em(i)} low-income housing; {\em(ii)} childcare services and disability care; and {\em(iii)} food assistance programs.

\paragraph{Low-Income Housing.}  As noted in the introduction, our research questions and model are motivated by the market for low-income housing.\footnote{Many redistribution programs in the low-income housing market have an income eligibility requirement.  We interpret the consumers in our model as the population of individuals who satisfy this eligibility requirement.}  In many of these markets, even among those eligible for public housing, there is substantial residual heterogeneity in both willingness to pay for quality (such as apartment size) and socioeconomic situation.\footnote{For example, \cite{waldinger21} and \cite{vandijk19} document significant heterogeneities among individuals eligible for public housing in Cambridge, Massachusetts, and Amsterdam, respectively.}  These factors are likely negatively correlated: for instance, a full-time MBA student and a minimum-wage worker may have similar current incomes, but the MBA student, with higher future earning potential, is expected to be more willing and able to rent a larger apartment.  

On one hand, our findings provide some support for policies that are currently in place.  In many countries, housing assistance programs represent a significant public expenditure \citep{curriegahvari08}.  When the correlation between willingness to pay for housing quality and socioeconomic status is stronger among eligible consumers (as reflected in the welfare weights) than in other markets (as captured by the opportunity cost of budget), \Cref{thm:scope} justifies housing assistance.  Housing assistance programs typically include a mix of instruments, such as public housing developments and nonlinear subsidies like the Low-Income Housing Tax Credit (LIHTC) program in the United States.  These instruments---whether by design or circumstance---target different segments of the population: for example, LIHTC tenants have higher average incomes than those who live in public housing developments \citep{collinsonetal16}.  While identifying the exact optimal mechanism is likely complex in practice, our \Cref{thm:optimal_negative} indicates that this tiered structure of housing assistance programs is sensible: consumers with lower willingness to pay (\eg, minimum-wage workers) should consume a public option or participate in a subsidy program, while those with higher willingness to pay (\eg, MBA students) should participate in a subsidy program or turn to the private market.  The size of housing assistance programs also varies widely from country to country; in Singapore, for example, 78.7\% of the population resides in some type of government-subsidized housing \citep{departmentofstatisticssingapore21}.  As our \Cref{prop:comp_statics} suggests, at least some of this variation may stem from different redistributive preferences and budget constraints across cities or countries.

On the other hand, our findings also demonstrate how current policies might be inadequate.  For example, public housing in the United States is often subject to rationing, as evidenced by long waitlists for public housing developments.  Our \Cref{thm:optimal_negative} indicates that, at least when the social planner can contract costlessly with private producers, rationing through allocation probability can {\em never} be optimal: the social planner can achieve better outcomes by implementing a deterministic allocation and adjusting the quality of housing provided.\footnote{However, as \cite{kang23} shows, rationing through allocation probability can be optimal when the social planner is less efficient than the private market (\eg, she can contract on only one quality level to be publicly provided).}  In addition, our \Cref{prop:non-market} provides theoretical support for universal access to housing---which has been advocated in policy discussions \citep{sitaramanalstott19}---under specific conditions on consumer welfare weights and the opportunity cost of budget.


\paragraph{Childcare Services and Disability Care.}  Our model and results can also be applied to the markets for childcare and disability care, albeit with some key distinctions. First, rather than interpreting $q$ as the quality of care, we treat it as the {\em quantity} of care consumed.  Second, instead of wealth or income, we interpret $\omega$ more broadly to reflect {\em social priorities}, such as placing greater weight on individuals with more children or those with more severe disabilities.

Given that the demand for care likely increases with these social priorities, we expect our results for the case of positive correlation to hold.  Specifically, our \Cref{thm:optimal_positive} supports subsidizing all individuals with sufficiently high demand for care.  If society places a sufficiently high value on redistributing to these individuals, or if other forms of redistribution are less effective---such as in cases where lump-sum cash transfers might lead to strategic behavior, like healthy individuals pretending to have disabilities---our \Cref{prop:non-market} provides justification for universal (free) access to a baseline level of care, which has been advocated in policy proposals \citep{sitaramanalstott19}.  While our model assumes that individuals are not allowed to supplement their care through the private market, this restriction is not binding in practice; as our \Cref{prop:topping} indicates, this constraint is not a limitation in the optimal program.

\paragraph{Food Assistance Programs.}  Finally, our model has limited applicability to food assistance programs.  On one hand, our model and results apply when demand for subsidized food is positively correlated with welfare weight.  For instance, in developing countries, governments often subsidize basic staples like coarse bread and cassava, which are inferior goods \citep{mackenzie91}.  In these cases, \Cref{prop:topping} suggests that the optimal redistribution program is nonetheless characterized by \Cref{thm:optimal_positive}: although consumers have the option to supplement consumption in the private market, they do not do so under the optimal mechanism.  On the other hand, our model and results do {\em not} apply when demand for the food product is negatively correlated with welfare weight.  For example, food voucher programs in developed countries, such as the Supplemental Nutrition Assistance Program in the United States, often allow---if not expect---participants to top up their food consumption in the private market.  In these cases, willingness to pay for everyday groceries is likely negatively correlated with socioeconomic status, even among eligible individuals. While the results in this paper do not directly address such cases, we provide a full analysis of the optimal redistribution program in a companion paper \citep{kangwatt24a}.

\section{Concluding Remarks}
\label{sec:conclusion}

The ability of consumers to access the private market is a prominent feature of many real-world redistribution programs, and---as we have shown in this paper---its inclusion in analyses affects the design of optimal mechanisms in substantial and realistic ways.  By developing a mechanism design model of in-kind redistribution that incorporates these participation constraints, we have quantified when a social planner can strictly improve on the laissez-faire outcome and characterized the optimal redistribution program as a combination of a free public option, nonlinear subsidies, and the laissez-faire allocation.  Our results highlight that while private market access limits the scope of in-kind redistribution, it also strengthens the case for non-market allocations, such as free public options.

While we have focused on private market access as the source of these participation constraints, the same constraints can also be interpreted as a Pareto improvement requirement.  Specifically, such a requirement arises when a social planner with full control over the entire market designs a mechanism that benefits all consumers relative to the laissez-faire outcome---an idea that naturally fits within political economy models of reform with consensus or majority voting, as noted by \cite{fuchsskrzypacz15}.  Recently, \cite{baronetal24} consider such a constraint in the reform of assignment mechanisms used to allocate Child Protective Services investigators.  Extending our approach in this paper to such models offers a promising avenue for future research.

Finally, although we have focused on redistribution as the social planner's motivation for intervention, our approach extends to mechanism design problems in other economic applications.  For example, externalities and paternalism can be accommodated in the social planner's objective (cf.~\citealp{akbarpouretal24b}; \citealp{kang24a}; \citealp{paistrack24}).  Aside from public programs, our approach can also be applied to nonlinear contracting settings, in which a profit-maximizing dominant firm competes with a competitive fringe (cf.~\citealp{kangmuir22}).  In such settings, the ability of the dominant firm to write exclusive contracts \citep{calzolaridenicolo13,calzolaridenicolo15} is mathematically equivalent to the ability of the social planner to prevent consumers from topping up their public consumption in the private market.  In ongoing work, we extend our approach to this setting to quantify the anticompetitive harm of exclusive contracting.  More generally, we believe that partial mechanism design, where only part of the market can be designed, represents a fruitful area for future work.

\clearpage

\bibliography{master_bibliography}
\bibliographystyle{econ-econometrica}

\clearpage

\renewcommand{\thesection}{Appendix A} 
\section{Derivation of Main Results}
\label{app:proof_main}
\renewcommand{\thesection}{A}
\renewcommand{\theequation}{\thesection.\arabic{equation}}
\setcounter{equation}{0}

In this appendix, we derive our main results, \Cref{thm:scope,thm:optimal_negative,thm:optimal_positive}.  To this end, we solve the social planner's problem stated at the end of \Cref{sec:model}:
\begin{align*}
    \max_{(q,t)}&\int_{\und\theta}^{\BAR\theta}\bigg[\omega(\theta)\underbrace{\left[\theta v(q(\theta))-t(\theta)\right]}_{\text{consumer surplus}} + \a\underbrace{\left[t(\theta)-cq(\theta)\right]}_{\text{total profit}}\bigg]\ \dd F(\theta)\\
    \text{s.t. }&(q,t) \text{ satisfies \eqref{eq:IC}, \eqref{eq:IR}, and \eqref{eq:LS}.}
\end{align*}

\subsection{Preliminary Analysis}\label{app:preliminary}

We begin by characterizing the \eqref{eq:IC} and \eqref{eq:LS} constraints in order to rewrite the social planner's problem in a more tractable form.  As these characterizations are relatively well-known, we state them without proof.

\begin{claim}\label{clm:IC}
    A mechanism $(q,t)$ satisfies \eqref{eq:IC} if and only if $q$ is nondecreasing and
    \[\theta v(q(\theta)) - t(\theta) = \underbrace{\und\theta v(q(\und\theta)) - t(\und\theta)}_{=:\und U} + \int_{\und\theta}^\theta v(q(s))\ \dd s.\]
\end{claim}

\Cref{clm:IC} follows from standard arguments in the mechanism design literature.  \citepos{myerson81} lemma implies that, given an allocation function $q$, there exists a payment function $t$ such that $(q,t)$ is incentive-compatible if and only if $q$ is nondecreasing.  The envelope theorem of \cite{milgromsegal02} uniquely identifies what the payment function must be, up to an additive constant, $\und U$, equal to the utility that the lowest consumer type receives under the mechanism.

\begin{claim}\label{clm:LS}
    An incentive-compatible mechanism $(q,t)$ satisfies \eqref{eq:LS} if and only if $\und U \leq \und\theta v(q(\und\theta))$.
\end{claim}

\Cref{clm:LS} shows that the social planner's restriction to mechanisms without lump-sum transfers is equivalent to imposing an upper bound on $\und U$ in terms of the quality $q(\und\theta)$ allocated to the lowest consumer type.  When \eqref{eq:LS} binds, $t(\und\theta)=0$; that is, the social planner allocates the good for free to the lowest consumer type.

We now reformulate the social planner's problem in utility space rather than allocation space.  To this end, we make the change of variables $\nu\coloneq v\circ q$; we refer  to $\nu$ henceforth as the subutility function induced by the mechanism.  Since $v$ is increasing by assumption, \Cref{clm:IC} implies that any incentive-compatible mechanism induces a nondecreasing subutility function. 

Next, we apply \Cref{clm:IC} to rewrite the \eqref{eq:IR} constraints and the social planner's objective.  Let $\und U^{\LF}$ be the utility that the lowest consumer type receives under the laissez-faire mechanism, and let $\nu^{\LF}$ be the subutility function induced by the laissez-faire mechanism.  Then the envelope theorem (cf.~\Cref{clm:IC}) implies that the \eqref{eq:IR} constraints can be written as
\[\und U + \int_{\und\theta}^\theta \nu(s)\ \dd s \geq \und U^{\LF} + \int_{\und\theta}^{\theta} \nu^{\LF}(s)\ \dd s\qquad\text{for any }\theta\in[\und\theta,\BAR\theta].\]
Moreover, for notational convenience, extend the domain of $v^{-1}$ to $\R$ and its range to the extended real line $\BAR\R\coloneq \R\cup\{+\infty\}$ by defining
\[\Psi(\hat\nu)\coloneq \begin{dcases}
    v^{-1}(\hat\nu) &\text{if }\hat\nu\in[v(0),v(A)],\\
    +\infty &\text{otherwise}.
\end{dcases}\]
Then the envelope theorem also allows us to rewrite the social planner's objective by eliminating dependence on the payment function: 
\[\eqref{eq:OBJ} = \left[\E[\omega]-\a\right]\und U +\int_{\und\theta}^{\BAR\theta}\left[\left[\a\theta-\frac{\int_\theta^{\BAR\theta}\left[\a-\omega(s)\right]\ \dd F(s)}{f(\theta)}\right]\nu(\theta)-\a c \Psi(\nu(\theta))\right]\ \dd F(\theta).\]

We summarize the above analysis by rewriting the social planner's problem.  Denote the set of nondecreasing functions by $\calI\coloneq\left\{h:[\und\theta,\BAR\theta]\to\R\text{ is nondecreasing}\right\}$.
Then \Cref{clm:IC,clm:LS} allow us to rewrite the social planner's problem as follows:
\begin{align*}
    \max_{\und U\in\R,\,\nu\in\calI}&\left\{\left[\E[\omega]-\a\right]\und U +\int_{\und\theta}^{\BAR\theta}\left[\left[\a\theta-\frac{\int_\theta^{\BAR\theta}\left[\a-\omega(s)\right]\ \dd F(s)}{f(\theta)}\right]\nu(\theta)-\a c \Psi(\nu(\theta))\right]\ \dd F(\theta)\right\}\\
    \text{s.t. }&\begin{dcases}
        \und U \leq \und\theta\nu(\und\theta),\\
        \und U +\int_{\und\theta}^\theta \nu(s)\ \dd s \geq \und U^{\LF} + \int_{\und\theta}^{\theta} \nu^{\LF} (s)\ \dd s\qquad\text{for any }\theta\in[\und\theta,\BAR\theta].
    \end{dcases}
\end{align*}

We conclude this subsection by making four observations about the social planner's problem that are clearer from this rewriting.

\begin{enumerate}
    \item {\bf Existence of solution.}  It is relatively straightforward to see that an optimal solution to the social planner's problem exists.  To see this, we endow $\calI$ with the $L^1$ topology and observe that the social planner's objective is continuous in $(\und U,\nu)$.  Without loss of generality, we focus on the set $K\coloneq\left\{h\in\calI:[\und\theta,\BAR\theta]\to[v(0),v(A)]\right\}$.  By the Helly selection theorem and the dominated convergence theorem, $K$ is compact; hence the constrained set (as a closed subset of $K$) is also compact.  Finally, $(\und U^{\LF},\nu^{\LF})\in K$ satisfies the constraints; hence the constrained set is nonempty.  We thus conclude that an optimal solution exists. 
    
    \item {\bf General uniqueness of solution.}  We now argue that the optimal solution to the social planner's problem is unique when $\E[\omega]\ne\a$.  Suppose on the contrary that $(\und U_1,\nu_1)$ and $(\und U_2,\nu_2)$ are distinct optimal solutions with distinct allocation functions $\nu_1\ne\nu_2$, and consider $(\und U^*,\nu^*) = \((\und U_1 + \und U_2)/2,(\nu_1 + \nu_2)/2\)$.  Clearly, $\nu^*$ is nondecreasing, and $(\und U^*,\nu^*)$ satisfies the constraints (since the constraints are linear).  However, $v^{-1}$ is strictly convex since $v$ is increasing and strictly concave by assumption; hence Jensen's inequality implies that the social planner's objective is strictly larger under $(\und U^*,\nu^*)$, contradicting the optimality of $(\und U_1,\nu_1)$ and $(\und U_2,\nu_2)$.
    
    The above argument shows that the optimal allocation function must be unique.  When the \eqref{eq:LS} constraint binds and/or when the \eqref{eq:IR} constraint binds for any type, $\und U^*$ is uniquely pinned down by the envelope theorem (cf.~\Cref{clm:IC}).  Thus the optimal solution to the social planner's problem can fail to be unique only if both the \eqref{eq:LS} and \eqref{eq:IR} constraints are slack, in which case $\E[\omega]=\a$.
    
    \item {\bf Optimality of deterministic mechanisms.}  Next, we observe that the social planner's restriction to deterministic mechanisms entails no loss of generality.  This follows from a similar argument to that used above for uniqueness: For any given stochastic mechanism $\chi$ (\ie, a probability distribution over deterministic mechanisms), we consider the deterministic mechanism $(\und U^*,\nu^*)$ corresponding to the arithmetic average of all deterministic mechanisms in the support of $\chi$.  Jensen's inequality then implies that the social planner's objective is weakly larger under $(\und U^*,\nu^*)$ than under $\chi$.
    
    \item {\bf Convex program with majorization constraints.}  Finally, we point out that the social planner's problem can be written as a convex program with majorization constraints.  To illustrate, suppose that the \eqref{eq:IR} constraint binds for the highest type: $\BAR U = \BAR U^{\LF}$.  (As we show in \Cref{lem:active_negative_1}, this holds when the private market is active and when welfare weight is negatively correlated with willingness to pay.)  Then the \eqref{eq:IR} constraints can be rewritten as
    \[\int_{\theta}^{\BAR\theta} \nu(s)\ \dd s \geq \int_{\theta}^{\BAR\theta} \nu^{\LF} (s)\ \dd s\qquad\text{for any }\theta\in[\und\theta,\BAR\theta].\]
    Since $\nu$ and $\nu^{\LF}$ are nondecreasing, this condition is equivalent to $\nu$ weakly majorizing $\nu^{\LF}$.
    This argument can be adapted for any type $\hat\theta$ whose \eqref{eq:IR} constraint binds by partitioning the type space into two intervals, $[\und\theta,\hat\theta]$ and $[\hat\theta,\BAR\theta]$: the social planner's problem on each segment is a convex program with majorization constraints.
    
    Recent papers (\eg, \citealp{kleineretal21}; \citealp{akbarpouretal24}) solve linear programs with majorization constraints and show that their solutions coincide with extreme points of the constrained set.  However, extreme point methods do not apply in our setting as our program is strictly convex (as $v$ is strictly concave since consumers have diminishing marginal utility); hence our solution is at an interior point.  As such, we have to develop an alternate approach to solving our problem.  We note that our approach provides an alternate way of recovering solutions to linear programs with majorization constraints by approximating any linear objective with strictly convex functions.
\end{enumerate}

\subsection{Full Mechanism Design}\label{app:full}

In this subsection, to convey intuition about the social planner's problem, we solve a relaxation of the social planner's problem where the \eqref{eq:IR} constraints are ignored and the lower bound $\und U\geq 0$ is instead imposed.  This relaxation corresponds to the assumption that the social planner has full control over the market (\ie, consumers can be prevented from consuming in the private market). 

\subsubsection{Case \texorpdfstring{\#}{}1: \texorpdfstring{$\E[\omega]\leq \a$}{E[ω]≤α}}\label{sec:full_weak_analysis}

We begin by supposing that $\E[\omega]\leq \a$.  Since $\a$ represents the social planner's opportunity cost of budget, this case captures settings where the social planner has other redistribution programs competing for the same budget, such as cash transfers that happen outside of the mechanism.

We now use standard tools from the mechanism design literature to solve the social planner's problem.  Since $\E[\omega]\leq\a$, the social planner can do no better than to choose the lowest possible value for $\und U$: $\und U^* =0$.  Consequently, the social planner's problem can be further simplified:
\[\max_{\nu\in\calI}\int_{\und\theta}^{\BAR\theta}\Bigg[\underbrace{\left[\a\theta-\frac{\int_\theta^{\BAR\theta}\left[\a-\omega(s)\right]\ \dd F(s)}{f(\theta)}\right]}_{\eqcolon \a J(\theta)}\nu(\theta) - \a c\Psi(\nu(\theta))\Bigg]\ \dd F(\theta).\]
In the above expression, $J$ can be thought of as a ``virtual welfare function'': it is the analog in our problem to the virtual valuation function (which obtains when $\omega\equiv0$) in mechanism design problems where the social planner's objective is to maximize profit.  

To solve this problem, let $\psi$ denote the derivative of $\Psi$.  Then the optimal subutility function in the solution to the social planner's problem is (cf.~\citealp{toikka11})
\[\nu^*(\theta) = \psi^{-1}\(\frac{\BAR J(\theta)}{c}\).\]

\paragraph{Negative correlation.}  To interpret this solution, we first consider the case where welfare weight is negatively correlated with willingness to pay---that is, $\omega$ is decreasing.  The solution to the social planner's problem in this case has two notable features.  First, the good is never offered to consumers for free, since the \eqref{eq:LS} constraint never binds.  Second, the social planner optimally distorts quality levels downwards from the laissez-faire quality levels.  To see this, observe that
\[J(\theta) = \theta - \frac{\int_\theta^{\BAR\theta}\left[\a-\omega(s)\right]\ \dd F(s)}{\a f(\theta)}\leq  \theta\qquad\text{because }\int_\theta^{\BAR\theta}\left[\a-\omega(s)\right]\ \dd F(s)\geq 0.\]
In turn, this implies that
\[\BAR J(\theta) \leq \theta\implies \nu^*(\theta) = \psi^{-1}\(\frac{\BAR J(\theta)}{c}\)\leq \psi^{-1}\(\frac{\theta}{c}\) = \nu^{\LF}(\theta).\]

Intuitively, these two features arise from the fact that the social planner has other redistribution programs available to her that are more efficient on the margin---as captured by the higher opportunity cost of budget, $\a \geq \E[\omega]$.  Consequently, she would prefer to redistribute with those programs instead of the good.  When she has full control of the market, she uses the good to tax consumers; by using quality to screen consumers, she can set different marginal tax rates for different consumers.

\paragraph{Positive correlation.}  We next consider the case where welfare weight is positively correlated with willingness to pay---that is, $\omega$ is increasing.  Again, the solution to the social planner's problem has two notable features.  First, as before, the good is never offered to consumers for free, since the \eqref{eq:LS} constraint never binds.  Second, the social planner optimally distorts quality levels upwards for consumers with higher types and downwards for consumers with lower types.  This is because, when $\omega$ is increasing, $\theta\mapsto\int_\theta^{\BAR\theta}\left[\a-\omega(s)\right]\ \dd F(s)$ is quasiconvex.  In turn, this implies that there exists $\hat\theta\in[\und\theta,\BAR\theta]$ such that
\[\BAR J(\theta)\begin{dcases} \geq \theta &\text{for }\theta\geq \hat\theta,\\
\leq \theta &\text{for }\theta\leq \hat\theta\end{dcases}
\implies \nu^*(\theta) \begin{dcases}
\geq \nu^{\LF}(\theta) &\text{for }\theta\geq\hat\theta,\\
\leq \nu^{\LF}(\theta) &\text{for }\theta\leq\hat\theta.\end{dcases}\]

Intuitively, these two features arise from the fact that the positive correlation between welfare weight and willingness to pay helps the social planner to target her redistribution program at consumers with higher welfare weights.  Consequently, even though the social planner has other redistribution programs available to her that are more efficient on the margin, she might still wish to subsidize the good for some consumers while taxing the good for other consumers.

\subsubsection{Case \texorpdfstring{\#}{}2: \texorpdfstring{$\E[\omega]> \a$}{E[ω]>α}}\label{sec:full_strong_analysis}

We now consider the case where $\E[\omega]>\a$.  This case captures settings where the social planner's outside options are less efficient, such as when there are political considerations that restrict the social planner from using cash transfers.

We again use standard tools from the mechanism design literature to solve the social planner's problem.  Since $\E[\omega]>\a$, the \eqref{eq:LS} constraint now binds; standard arguments imply that the optimal Lagrange multiplier for the \eqref{eq:LS} constraint must be $\mu^*=\E[\omega]-\a>0$.  Consequently, the social planner's optimal subutility function $\nu^*$ solves
\[\max_{\nu\in\calI}\int_{\und\theta}^{\BAR\theta}\left[\left[\a J(\theta) + \frac{\E[\omega] - \a}{f(\theta)}\cdot\und\theta\d_{\und\theta}(\theta)\right]\nu(\theta) - \a c\Psi(\nu(\theta))\right]\ \dd F(\theta).\]
Thus we deduce that
\[\nu^*(\theta) = \psi^{-1}\(\frac1{c}\cdot\BAR{\(s\mapsto J(s) + \frac{\E[\omega]-\a}{\a f(s)}\cdot\und\theta\d_{\und\theta}(s)\)}(\theta)\).\]

\paragraph{Negative correlation.} Suppose that welfare weight is negatively correlated with willingness to pay.  As before, we point out two notable features of the solution to the social planner's problem.  First, the lowest quality of the good, $\nu^*(\und\theta)$, is now made available for free to a positive mass of consumers with the lowest types.  Equivalently, the social planner optimally imposes a (binding) minimum quality level for goods sold in the market in order to ensure that the poorest consumers are allocated a higher quality level than what they would otherwise consume.  Second, the social planner optimally distorts quality levels upwards for low-type consumers---and downwards for high-type consumers---from the laissez-faire quality levels.  Under additional assumptions about the social welfare weight function (\eg, $\omega(\theta)>\a$ for any $\theta\in[\und\theta,\BAR\theta]$), the social planner optimally distorts quality levels upwards for {\em all} consumers.

Intuitively, these two features arise from the fact that the good is more efficient on the margin than other redistribution programs available to the social planner.  Consequently, she uses the good to redistribute to low-type consumers, while raising revenue from taxes outside of the mechanism.  Depending on the social welfare weights, the social planner might also tax higher quality levels of the good that high-type consumers consume.

\paragraph{Positive correlation.}  Finally, consider the case where welfare weight is positively correlated with willingness to pay.  Similar to the case of negative correlation, the lowest quality of the good is made available for free to a positive mass of consumers with the lowest types.  The social planner optimally distorts upwards for {\em all} consumers relative to the laissez-faire quality levels.

Because the good is more efficient on the margin than other redistribution programs available to the social planner, the social planner redistributes with the good.  The positive correlation between welfare weight and willingness to pay also helps the social planner to target her redistribution program at consumers with higher welfare weights.  As such, the social planner subsidizes the good to benefit all consumers and raises revenue from taxes outside of the mechanism.


\subsection{Partial Mechanism Design: Negative Correlation}\label{sec:negative_proof}

We now solve the social planner's problem with the original \eqref{eq:IR} constraints, assuming that welfare weight is negatively correlated with willingness to pay (\ie, $\omega$ is decreasing).

\subsubsection{Case \texorpdfstring{\#}{}1: \texorpdfstring{$\E[\omega]\leq \a$}{E[ω]≤α}}

We begin by supposing that $\E[\omega]\leq\a$.  Following the approach of \cite{amadorbagwell13}, we solve the social planner's problem in three steps: {\em(i)} guessing Lagrange multipliers; {\em(ii)} maximizing the Lagrangian; and {\em(iii)} applying the \cite{luenberger69} sufficiency theorem.

\subsubsection*{\ul{Guessing Lagrangian multipliers}.}

Let $\mu\in\R_+$ and the nondecreasing function $\La:[\und\theta,\BAR\theta]\to\R$ (where we normalize $\La(\BAR\theta)=\a-\E[\omega]$ without loss of generality) respectively denote the Lagrange multipliers for the \eqref{eq:LS} constraint and the \eqref{eq:IR} constraint, so that the Lagrangian for the social planner's problem can be written as
\begin{align*}
    \calL(\und U,\nu;\mu,\La)
    &= \left[\E[\omega]-\a-\mu\right]\und U +\int_{\und\theta}^{\BAR\theta}\left[\und U - \und U^{\LF} + \int_{\und\theta}^\theta\left[\nu(s)-\nu^{\LF}(s)\right]\ \dd s\right]\ \dd\La(\theta)\\
    &\qquad+ \int_{\und\theta}^{\BAR\theta}\left[\left[\a\theta-\frac{\int_\theta^{\BAR\theta}\left[\a-\omega(s)\right]\ \dd F(s)}{f(\theta)} + \frac{\mu\und\theta\d_{\und\theta}(\theta)}{f(\theta)}\right]\nu(\theta)-\a c\Psi(\nu(\theta))\right]\ \dd F(\theta)\\
    &=-\left[\mu + \La(\und\theta)\right]\und U - \int_{\und\theta}^{\BAR\theta}\left[\und U^{\LF} + \int_{\und\theta}^\theta\nu^{\LF}(s)\ \dd s\right]\ \dd\La(\theta)\\
    &\qquad+ \int_{\und\theta}^{\BAR\theta}\left[\left[\a\theta+\frac{\int_{\und\theta}^{\theta}\left[\a-\omega(s)\right]\ \dd F(s)+\mu\und\theta\d_{\und\theta}(\theta) -\La(\theta)}{f(\theta)}\right]\nu(\theta)-\a c\Psi(\nu(\theta))\right]\ \dd F(\theta).
\end{align*}
Let $\mu_{\max}\coloneq -\min_{\theta\in[\und\theta,\BAR\theta]}\int_{\und\theta}^\theta\left[\a-\omega(s)\right]\ \dd F(s)$.  For each $\mu\in[0,\mu_{\max}]$, define the function
\[\theta_H(\mu) \coloneq \max\left\{\theta\in[\und\theta,\BAR\theta]:\int_{\und\theta}^\theta\left[\a-\omega(s)\right]\ \dd F(s) \leq -\mu\right\}.\]
Also, for each $\mu\in\R$, define 
\begin{equation}\label{eq:nu_mu}
\nu_\mu(\theta)\coloneq\psi^{-1}\(\BAR{\left.\(s\mapsto\frac{s}{c} + \frac{\int_{\und\theta}^s\left[\a-\omega(z)\right]\ \dd F(z) + \mu\und\theta\d_{\und\theta}(s)+\mu}{\a cf(s)}\)\right|_{[\und\theta,\theta_H(\mu)]}}\)(\theta).
\end{equation}
We guess the optimal Lagrange multiplier for the \eqref{eq:LS} constraint to be
\begin{equation}\label{eq:mu}
    \mu^* \coloneq \min\left\{\mu\in[0,\mu_{\max}]:\int_{\und\theta}^{\theta_H(\mu)}\nu_{\mu}(s)\ \dd s + \und\theta\nu_{\mu}(\und\theta)- U^{\LF}(\theta_H(\mu))\geq0\right\}.
\end{equation}
Observe that $\mu^*$ is well-defined because {\em(i)}~the left-hand side of the inequality in \cref{eq:mu} is continuous in $\mu$; and {\em(ii)}~the inequality holds when $\mu=\mu_{\max}$.  Here, observation {\em(ii)} follows from the fact that $\nu_{\mu_{\max}} \geq \nu^{\LF}$ pointwise by construction, which implies that
\begin{align*}
\int_{\und\theta}^{\theta_H(\mu_{\max})}\nu_{\mu_{\max}}(s)\ \dd s + \und\theta \nu_{\mu_{\max}}(\und\theta)
&\geq \int_{\und\theta}^{\theta_H(\mu_{\max})}\nu^{\LF}(s)\ \dd s + \und\theta \nu^{\LF}(\und\theta)\\
&\geq \int_{\und\theta}^{\theta_H(\mu_{\max})}\nu^{\LF}(s)\ \dd s +  \und U^{\LF} = U^{\LF}(\theta_H(\mu_{\max})).
\end{align*}
Below, we denote $\theta_H^*\coloneq \theta_H(\mu^*)$.  
In addition, we guess the optimal Lagrange multiplier for the \eqref{eq:IR} constraint to be
\begin{equation}\label{eq:La}
    \La^*(\theta)\coloneq
    \begin{dcases}
        \int_{\und\theta}^{\theta}\left[\a-\omega(s)\right]\ \dd F(s) &\text{if }\theta> \theta_H^*,\\
        -\mu^* &\text{otherwise}.
    \end{dcases}
\end{equation}
Note that  $\La^*$ is nondecreasing by construction.  This is because $\theta\mapsto\int_{\und\theta}^\theta\left[\a-\omega(s)\right]\ \dd F(s)$ is quasiconvex: 
\[\frac{\pp}{\pp\theta}\int_{\und\theta}^{\theta}\left[\a-\omega(s)\right]\ \dd F(s) = \left[\a-\omega(\theta)\right]f(\theta),\] which crosses zero at most once since $\omega$ is decreasing.

\subsubsection*{\ul{Maximizing the Lagrangian}.}

Given these guesses for $\mu^*$ and $\La^*$, we set
\[\und U^* \coloneq \begin{dcases}
    U^{\LF}(\theta_H^*) - \int_{\und\theta}^{\theta_H^*}\nu^*(s)\ \dd s &\text{if }\mu^*=0,\\
    \und\theta\nu^*(\und\theta) &\text{if }\mu^*>0.
\end{dcases}\]
Moreover, we define the subutility function $\nu^*$ as follows:
\[\nu^*(\theta) =\begin{dcases}
   \nu^{\LF}(\theta) &\text{if }\theta\geq \theta_H^*,\\
   \nu_{\mu^*}(\theta) &\text{otherwise}.
\end{dcases}\]
We now demonstrate that $\nu^*$ is nondecreasing and maximizes the Lagrangian given the Lagrange multipliers $\mu^*$ and $\La^*$.  
\begin{enumerate}[label=({\bf\arabic*})]
\item {\bf $\nu^*$ is nondecreasing.}

It suffices to verify that $\nu_{\mu^*}(\theta_H^*)\leq \nu^{\LF}(\theta_H^*)$.  If $\mu^*=0$, then observe that
\[\int_{\und\theta}^\theta\left[\a-\omega(s)\right]\ \dd F(s) \leq 0\implies \frac{\theta}{c} + \frac{\int_{\und\theta}^\theta\left[\a-\omega(s)\right]\ \dd F(s)}{\a cf(\theta)}\leq \frac{\theta}{c}\qquad\text{for any }\theta<\theta_H^*.\]
In turn, this implies that $\psi(\nu_{\mu^*}(\theta))\leq \theta/c\leq \theta_H^*/c=\psi(\nu^{\LF}(\theta_H^*))$ for $\theta\leq \theta_H^*$.

If $\mu^*>0$, suppose on the contrary that $\nu_{\mu^*}(\theta_H^*)>\nu^{\LF}(\theta_H^*)$.  This can happen only if $\theta_H^*$ lies in a pooling region on $\nu_{\mu^*}$; let this pooling region be denoted by $[\theta_-,\theta_H^*]\subset[\und\theta,\theta_H^*]$.  On one hand, if $\theta_-=\und\theta$, then
\begin{align*}
\int_{\und\theta}^{\theta_H^*}\nu_{\mu^*}(s)\ \dd s + \und\theta \nu_{\mu^*}(\und\theta)
&=\int_{\und\theta}^{\theta_H^*}\nu_{\mu^*}(\theta_H^*)\ \dd s + \und\theta \nu_{\mu^*}(\theta_H^*) = \theta_H^*\nu_{\mu^*}(\theta_H^*) > U^{\LF}(\theta_H^*).
\end{align*}
However, recall from the construction of $\mu^*$ that, because the left-hand side of inequality in \cref{eq:mu} is continuous in $\mu$ and $\mu^*>0$,
\[U^{\LF}(\theta_H^*) = \int_{\und\theta}^{\theta_H^*}\nu_{\mu^*}(s)\ \dd s + \und\theta\nu_{\mu^*}(\und\theta),\qquad\text{a contradiction}.\]
On the other hand, if $\theta_-\ne\und\theta$, then 
\[\psi(\nu_{\mu^*}(\theta_-)) = \frac{\theta_-}{c} + \frac{\int_{\und\theta}^{\theta_-}\left[\a-\omega(s)\right]\ \dd F(s) + \mu^*}{\a cf(\theta_-)} = \psi(\nu_{\mu^*}(\theta_H^*)) > \psi(\nu^{\LF}(\theta_H^*)) = \frac{\theta_H^*}{c}.\]
In turn, this implies that $\int_{\und\theta}^{\theta_-}\left[\a-\omega(s)\right]\ \dd F(s) + \mu^* \geq 0$.  Since $\theta\mapsto \int_{\und\theta}^\theta\left[\a-\omega(s)\right]\ \dd F(s)$ is quasiconvex, it follows that
\[ \int_{\und\theta}^\theta\left[\a-\omega(s)\right]\ \dd F(s) + \mu^*\geq 0\qquad\text{for any }\theta\in[\und\theta,\theta_-].\]
Consequently, $\nu_\mu^*(\theta)\geq \nu^{\LF}(\theta)$ for any $\theta\in[\und\theta,\theta_-]$; hence a similar argument as before yields
\begin{align*}
U^{\LF}(\theta_H^*) &= \int_{\und\theta}^{\theta_H^*}\nu_{\mu^*}(s)\ \dd s + \und\theta \nu_{\mu^*}(\und\theta)\\
&= \int_{\und\theta}^{\theta_-}\nu_{\mu^*}(s)\ \dd s + \int_{\theta_-}^{\theta_H^*}\nu_{\mu^*}(\theta_H^*)\ \dd s + \und\theta \nu_{\mu^*}(\und\theta)\geq U^{\LF}(\theta_-) + \int_{\theta_-}^{\theta_H^*}\nu_{\mu^*}(\theta_H^*)\ \dd s.
\end{align*}
But this yields the following contradiction:
\begin{align*}
U^{\LF}(\theta_H^*) - U^{\LF}(\theta_-) = \int_{\theta_-}^{\theta_H^*}\nu^{\LF}(s)\ \dd s&\geq \int_{\theta_-}^{\theta_H^*}\nu_{\mu^*}(\theta_H^*)\ \dd s > \int_{\theta_-}^{\theta_H^*}\nu^{\LF}(\theta_H^*)\ \dd s.
\end{align*}

\item {\bf $\nu^*$ maximizes the Lagrangian.}

It suffices to verify that, for any $\nu\in\calI$ satisfying the \eqref{eq:IR} constraint, the following variational inequality is satisfied (see, \eg, Chapter II of \citealp{ekelandtemam99}):
\[\int_{\und\theta}^{\BAR\theta}\left[\nu^*(\theta)-\nu(\theta)\right]\left[\frac{\theta}{c}+\frac{\int_{\und\theta}^{\theta}\left[\a-\omega(s)\right]\ \dd F(s)+\mu^*\und\theta\d_{\und\theta}(\theta) -\La^*(\theta)}{\a cf(\theta)} - \psi(\nu^*(\theta))\right]\ \dd F(\theta)\geq 0.\]
Clearly, by our construction of $\La^*$ in \cref{eq:La}, 
\begin{align*}
&\int_{\theta_H^*}^{\BAR\theta}\left[\nu^*(\theta)-\nu(\theta)\right]\underbrace{\left[\frac{\theta}{c}+\frac{\int_{\und\theta}^{\theta}\left[\a-\omega(s)\right]\ \dd F(s)+\mu^*\und\theta\d_{\und\theta}(\theta) -\La^*(\theta)}{\a cf(\theta)} - \psi(\nu^*(\theta))\right]}_{=0\text{ for }\theta\in[\theta_H^*,\BAR\theta]}\ \dd F(\theta) =0. 
\end{align*}
Consequently, it suffices to show that
\[\int_{\und\theta}^{\theta_H^*}\left[\nu^*(\theta)-\nu(\theta)\right]\left[\frac{\theta}{c} + \frac{\int_{\und\theta}^\theta\left[\a-\omega(s)\right]\ \dd F(s) + \mu^*\und\theta\d_{\und\theta}(\theta) + \mu^*}{\a cf(\theta)} - \psi(\nu^*(\theta))\right]\ \dd F(\theta)\geq 0.\]
However, this follows from our construction of $\nu_{\mu^*}$ in \cref{eq:nu_mu} (cf.~\citealp{toikka11}).
\end{enumerate}

Following the above discussion, we conclude that  $(\und U^*,\nu^*)$ maximizes the Lagrangian, given the Lagrange multipliers $\mu^*$ and $\La^*$:
\[(\und U^*,\nu^*)\in\argmax_{\und U\in\R_+,\,\nu\in\calI}\calL(\und U,\nu;\mu^*,\La^*).\]
Note that the results of \cite{toikka11} imply that the $\nu^*$ can be alternatively be written as
\begin{equation}\label{eq:neg}
    \nu^*(\theta) = \psi^{-1}\(\BAR{\(s\mapsto \frac{s}{c}+\frac{\int_{\und\theta}^{s}\left[\a-\omega(z)\right]\ \dd F(z)+\mu^*\und\theta\d_{\und\theta}(s) -\La^*(s)}{\a cf(s)}\)}(\theta)\).
\end{equation}

%


\subsubsection*{\ul{Applying the Luenberger sufficiency theorem}.}

We now apply the \citeauthor{luenberger69} sufficiency theorem.  To this end, observe that:
\begin{enumerate}[label=({\bf\arabic*})]
\item {\bf $(\und U^*,\nu^*)$ satisfies the \eqref{eq:LS}  constraint.}

If $\mu^*>0$, then $\und U^*=\und\theta\nu^*(\und\theta)$ by construction.  If $\mu^*=0$, then 
\[\und U^* =  U^{\LF}(\theta_H^*) - \int_{\und\theta}^{\theta_H^*}\nu^*(s)\ \dd s\leq \und\theta \nu^*(\und\theta),\]
where the inequality follows from the construction of $\mu^*$ in \cref{eq:mu}.  

\item {\bf $(\und U^*,\nu^*)$ satisfies the \eqref{eq:IR}  constraint.}

Observe that our construction of $\und U^*$ (for the case of $\mu^*=0$) and  \cref{eq:mu} (for the case of $\mu^*>0$) imply that
\[\und U^* + \int_{\und\theta}^{\theta_H^*}\nu^*(s)\ \dd s = U^{\LF}(\theta_H^*)=\und U^{\LF} + \int_{\und\theta}^{\theta_H^*}\nu^{\LF}(s)\ \dd s.\]
Since $\nu^*(\theta)=\nu^{\LF}(\theta)$ for $\theta\in[\theta_H^*,\BAR\theta]$, it follows that the \eqref{eq:IR} constraint holds for all $\theta\in[\theta_H^*,\BAR\theta]$.  

It remains to check that the \eqref{eq:IR} constraint holds for all $\theta\in[\und\theta,\theta_H^*]$.  To this end, observe that the following generalized function is quasiconvex:
\[\theta\mapsto \int_{\und\theta}^\theta\left[\a-\omega(s)\right]\ \dd F(s) + \mu^*\und\theta\d_{\und\theta}(\theta) +\mu^*.\]
This implies that $\nu^*(\theta)$ crosses $\nu^{\LF}(\theta)$ at most once (from above) for $\theta\in(\und\theta,\theta_H^*)$; hence it suffices to check that the \eqref{eq:IR} constraint holds for $\theta=\und\theta$.  On one hand, if $\mu^*>0$, then \cref{eq:mu} implies that
\[\und\theta \nu^*(\und\theta) + \int_{\und\theta}^{\theta_H^*} \nu^*(s)\ \dd s = U^{\LF}(\theta_H^*) = \und U^{\LF} + \int_{\und\theta}^{\theta_H^*}\nu^{\LF}(s)\ \dd s.\]
Rearranging yields
\[\int_{\und\theta}^{\theta_H^*}\left[\nu^{\LF}(s)-\nu^*(s)\right]\ \dd s = \und\theta \nu^*(\und\theta) - \und U^{\LF} \geq\und\theta \left[\nu^*(\und\theta)-\nu^{\LF}(\und\theta)\right] \geq 0.\]
In turn, this means that
\[\und U^* = U^{\LF}(\theta_H^*) -\int_{\und\theta}^{\theta_H^*}\nu^*(s)\ \dd s\geq U^{\LF}(\theta_H^*)-\int_{\und\theta}^{\theta_H^*}\nu^{\LF}(s)\ \dd s=\und U^{\LF}.\]
On the other hand, if $\mu^*=0$, then $\nu^*(\theta)\leq \nu^{\LF}(\theta)$ for any $\theta\in[\und\theta,\theta_H^*]$; hence
\[\und U^* = U^{\LF}(\theta_H^*) -\int_{\und\theta}^{\theta_H^*}\nu^*(s)\ \dd s\geq U^{\LF}(\theta_H^*)-\int_{\und\theta}^{\theta_H^*}\nu^{\LF}(s)\ \dd s=\und U^{\LF}.\]

\item {\bf The complementary slackness conditions are satisfied.}

It is clear that
\[\mu^*\left[\und\theta\nu^*(\und\theta)-\und U^*\right] = 0.\]
Moreover, observe that
\begin{align*}
&\int_{\und\theta}^{\BAR\theta}\left[\und U^* - \und U^{\LF} + \int_{\und\theta}^{\theta}\left[\nu^*(s)-\nu^{\LF}(s)\right]\ \dd s \right]\ \dd\La^*(\theta)\\
&=\int_{\theta_H^*}^{\BAR\theta}\left[\und U^* - \und U^{\LF} + \int_{\und\theta}^{\theta}\left[\nu^*(s)-\nu^{\LF}(s)\right]\ \dd s \right]\ \dd\La^*(\theta).
\end{align*}
Observe that our construction of $\und U^*$ (for the case of $\mu^*=0$) and  \cref{eq:mu} (for the case of $\mu^*>0$) imply that
\[\und U^* + \int_{\und\theta}^{\theta_H^*}\nu^*(s)\ \dd s = U^{\LF}(\theta_H^*)=\und U^{\LF} + \int_{\und\theta}^{\theta_H^*}\nu^{\LF}(s)\ \dd s.\]
Since $\nu^*(\theta)=\nu^{\LF}(\theta)$ for $\theta\in[\theta_H^*,\BAR\theta]$, it follows that the desired complementary slackness condition holds.
\end{enumerate}

\subsubsection{Case \texorpdfstring{\#}{}2: \texorpdfstring{$\E[\omega]> \a$}{E[ω]>α}}

We now consider the case where $\E[\omega] > \a$.  Instead of the social planner's problem, however, we study the following relaxation.
\begin{align*}
    \max_{\und U\in\R,\,\nu\in\calI}&\left\{\left[\E[\omega]-\a\right]\und U +\int_{\und\theta}^{\BAR\theta}\left[\left[\a\theta-\frac{\int_\theta^{\BAR\theta}\left[\a-\omega(s)\right]\ \dd F(s)}{f(\theta)}\right]\nu(\theta)-\a c \Psi(\nu(\theta))\right]\ \dd F(\theta)\right\}\\
    \text{s.t. }&\begin{dcases}
        \und U \leq \und\theta\nu(\und\theta),\\
        \und U +\int_{\und\theta}^{\BAR\theta} \nu(s)\ \dd s \geq \und U^{\LF} + \int_{\und\theta}^{\BAR\theta} \nu^{\LF} (s)\ \dd s.
    \end{dcases}
\end{align*}
This relaxation requires the \eqref{eq:IR} constraint to hold only for the highest consumer type, $\BAR\theta$, rather than for all consumer types.  We begin by solving this relaxed problem, and then we subsequently demonstrate that its solution also solves the social planner's problem.

First, suppose that the \eqref{eq:IR} constraint is slack.  In this case, the solution to the relaxed problem coincides with the solution to the full mechanism design problem studied in \Cref{sec:full_strong_analysis}:
\begin{equation}\label{eq:partial_strong_solution}
    \und U^* = \und\theta\nu^*(\und\theta)\AND\nu^*(\theta) = \psi^{-1}\(\BAR{\(s\mapsto \frac{J(s)}{c} + \frac{\E[\omega]-\a}{\a cf(s)}\cdot \und\theta\d_{\und\theta}(s)\)}(\theta)\).
\end{equation}
In order for the \eqref{eq:IR} constraint to be slack, we require that
\begin{equation}\label{eq:partial_strong_condition}
    \und U^* + \int_{\und\theta}^{\BAR\theta}\nu^*(s)\ \dd s > \und U^{\LF} + \int_{\und\theta}^{\BAR\theta}\nu^{\LF}(s)\ \dd s.
\end{equation}
We now argue that, if this condition holds, then the solution must also solve the social planner's problem.  Indeed, suppose that the \eqref{eq:IR} constraint is violated for some type $\hat\theta\in[\und\theta,\BAR\theta]$.  Then $\hat\theta\ne\und\theta$ since $\und U^* \geq \und U^{\LF}$ by construction.  Moreover, since quality is distorted upwards for low types (cf.~\Cref{sec:full_strong_analysis}), the envelope theorem (cf.~\Cref{clm:IC}) implies that the \eqref{eq:IR} constraint is satisfied by these low types; hence quality is distorted downwards at $\hat\theta$.  Consequently, quality is distorted downwards for all types $\theta\in[\hat\theta,\BAR\theta]$; thus the envelope theorem implies that the \eqref{eq:IR} constraint must in fact also be violated for the highest type, $\BAR\theta$.  But this contradicts the condition \eqref{eq:partial_strong_condition}, which requires that the \eqref{eq:IR} constraint is satisfied for the highest type.  We conclude that, under the condition \eqref{eq:partial_strong_condition}, the solution to the social planner's problem is in fact given by \cref{eq:partial_strong_solution}.

Next, suppose that \eqref{eq:partial_strong_condition} fails to hold, so that the \eqref{eq:IR} constraint binds in the relaxed problem.  In this case, the solution to the relaxed problem no longer coincides with the solution to the full mechanism design problem studied in \Cref{sec:full_strong_analysis}.  However, we know that the relaxed problem nonetheless admits a solution since $(\und U^{\LF},\nu^{\LF})$ satisfies both the \eqref{eq:LS} constraint and the (relaxed) \eqref{eq:IR} constraint.  Denote the solution to the relaxed problem by $(\und U^*,\nu^*)$, and denote the optimal Lagrange multipliers for the \eqref{eq:LS} and relaxed \eqref{eq:IR} constraints by $\mu^*,\la^*\geq 0$ respectively.  Then $\mu^* - \la^* = \E[\omega] - \a$.  On one hand, if $\mu^*>0$, then the solution can be written as
\begin{equation}\label{eq:partial_strong_solution_2}
    \und U^* = \und\theta \nu^*(\und\theta)\AND \nu^*(\theta) = \psi^{-1}\(\BAR{\(s\mapsto \frac{J(s)}{c} + \frac{\a - \E[\omega] + \mu^* + \mu^*\und\theta\d_{\und\theta}(s)}{\a cf(s)}\)}(\theta)\).
\end{equation}
While this differs from \cref{eq:partial_strong_solution} in that $\la^*>0$ when the \eqref{eq:IR} constraint binds, a similar argument as before implies that the \eqref{eq:IR} constraint must in fact hold for every type.  On the other hand, if $\mu^*=0$, then note that $\la^* = \a-\E[\omega]<0$, a contradiction.  Consequently, the solution to the social planner's problem is in fact generally given by \cref{eq:partial_strong_solution_2}, regardless of whether the condition \eqref{eq:partial_strong_condition} holds.

\subsection{Partial Mechanism Design: Positive Correlation}\label{sec:positive_proof}

Finally, we solve the social planner's problem with the original \eqref{eq:IR} constraints, assuming that welfare weight is positively correlated with willingness to pay (\ie, $\omega$ is increasing).

\subsubsection{Case \texorpdfstring{\#}{}1: \texorpdfstring{$\E[\omega]\leq \a$}{E[ω]≤α}}

We begin by supposing that $\E[\omega]\leq\a$.  We follow our approach in Appendix~\ref{sec:negative_proof} by guessing Lagrangian multipliers, maximizing the Lagrangian, and then applying the \citeauthor{luenberger69} sufficiency theorem.

\subsubsection*{\ul{Guessing Lagrangian multipliers}.}

Let $\mu\in\R_+$ and the nondecreasing function $\La:[\und\theta,\BAR\theta]\to\R$ (where we normalize $\La(\und\theta)=0$ without loss of generality) respectively denote the Lagrange multipliers for the \eqref{eq:LS} constraint and the \eqref{eq:IR} constraint, so that the Lagrangian for the social planner's problem can be written as
\begin{align*}
    \calL(\und U,\nu;\mu,\La)
    &= \left[\E[\omega]-\a-\mu\right]\und U +\int_{\und\theta}^{\BAR\theta}\left[\und U - \und U^{\LF} + \int_{\und\theta}^\theta\left[\nu(s)-\nu^{\LF}(s)\right]\ \dd s\right]\ \dd\La(\theta)\\
    &\qquad+ \int_{\und\theta}^{\BAR\theta}\left[\left[\a\theta-\frac{\int_\theta^{\BAR\theta}\left[\a-\omega(s)\right]\ \dd F(s)}{f(\theta)} + \frac{\mu\und\theta\d_{\und\theta}(\theta)}{f(\theta)}\right]\nu(\theta)-\a c\Psi(\nu(\theta))\right]\ \dd F(\theta)\\
    &=\left[\E[\omega]-\a-\mu + \La(\BAR\theta)\right]\und U - \int_{\und\theta}^{\BAR\theta}\left[\und U^{\LF} + \int_{\und\theta}^\theta\nu^{\LF}(s)\ \dd s\right]\ \dd\La(\theta)\\
    &\qquad+ \int_{\und\theta}^{\BAR\theta}\left[\left[J(\theta) + \frac{\mu\und\theta\d_{\und\theta}(\theta) + \La(\BAR\theta) -\La(\theta)}{f(\theta)}\right]\nu(\theta)-\a c\Psi(\nu(\theta))\right]\ \dd F(\theta).
\end{align*}
Define
\[\theta_L^* \coloneq \min\left\{\theta\in[\und\theta,\BAR\theta]:\int_{\und\theta}^{\theta}\left[\a-\omega(s)\right]\ \dd F(s) \geq \a - \E[\omega]\right\}.\]
Also, define
\begin{equation}\label{eq:nu_mu_positive}
\nu_0(\theta)\coloneq\psi^{-1}\(\BAR{\left.\(s\mapsto\frac{s}{c} - \frac{\int_{s}^{\BAR\theta}\left[\a-\omega(z)\right]\ \dd F(z)}{\a cf(s)}\)\right|_{[\theta_L^*,\BAR\theta]}}\)(\theta).
\end{equation}
We guess the Lagrange multipliers $\mu^*=0$ and
\begin{equation}\label{eq:La_positive}
\La^*(\theta) \coloneq \begin{dcases}
    \a-\E[\omega] &\text{if }\theta > \theta_L^*,\\
    \int_{\und\theta}^{\theta}\left[\a-\omega(s)\right]\ \dd F(s) &\text{if }\theta\leq\theta_L^*.
\end{dcases}
\end{equation}
Note that  $\La^*$ is nondecreasing by construction.  This is because $\theta\mapsto\int_{\und\theta}^\theta\left[\a-\omega(s)\right]\ \dd F(s)$ is quasiconcave: 
\[\frac{\pp}{\pp\theta}\int_{\und\theta}^{\theta}\left[\a-\omega(s)\right]\ \dd F(s) = \left[\a-\omega(\theta)\right]f(\theta),\] which crosses zero at most once since $\omega$ is increasing.

\subsubsection*{\ul{Maximizing the Lagrangian}.}

Given these guesses for $\mu^*$ and $\La^*$, we set $\und U^* = \und U^{\LF}$. 
Moreover, we define the subutility function $\nu^*$ as follows:
\[\nu^*(\theta) =\begin{dcases}
   \nu_{0}(\theta) &\text{if }\theta\geq \theta_L^*,\\
   \nu^{\LF}(\theta) &\text{otherwise}.
\end{dcases}\]
We now demonstrate that $\nu^*$ is nondecreasing and maximizes the Lagrangian given the Lagrange multipliers $\mu^*$ and $\La^*$.  
\begin{enumerate}[label=({\bf\arabic*})]
\item {\bf $\nu^*$ is nondecreasing.}

It suffices to verify that $\nu_{0}(\theta_L^*)\geq \nu^{\LF}(\theta_L^*)$.  Indeed, observe that
\[\int_{\theta}^{\BAR\theta}\left[\a-\omega(s)\right]\ \dd F(s) \leq 0\implies \frac{\theta}{c} - \frac{\int_{\theta}^{\BAR\theta}\left[\a-\omega(s)\right]\ \dd F(s)}{\a cf(\theta)}\geq \frac{\theta}{c}\qquad\text{for any }\theta\geq\theta_L^*.\]
In turn, this implies that $\psi(\nu_{0}(\theta))\geq \theta/c\geq \theta_L^*/c=\psi(\nu^{\LF}(\theta_L^*))$ for $\theta\geq \theta_L^*$.

\item {\bf $\nu^*$ maximizes the Lagrangian.}

It suffices to verify that, for any $\nu\in\calI$ satisfying the \eqref{eq:IR} constraint, the following variational inequality is satisfied:
\[\int_{\und\theta}^{\BAR\theta}\left[\nu^*(\theta)-\nu(\theta)\right]\left[\frac{\theta}{c}+\frac{\La^*(\BAR\theta)-\La^*(\theta)-\int_{\theta}^{\BAR\theta}\left[\a-\omega(s)\right]\ \dd F(s)}{\a cf(\theta)} - \psi(\nu^*(\theta))\right]\ \dd F(\theta)\geq 0.\]
Clearly, by our construction of $\La^*$ in \cref{eq:La_positive}, 
\begin{align*}
&\int_{\und\theta}^{\theta_L^*}\left[\nu^*(\theta)-\nu(\theta)\right]\underbrace{\left[\frac{\theta}{c}+\frac{\int_{\und\theta}^{\theta}\left[\a-\omega(s)\right]\ \dd F(s)-\La^*(\theta)}{\a cf(\theta)} - \psi(\nu^*(\theta))\right]}_{=0\text{ for }\theta\in[\und\theta,\theta_L^*]}\ \dd F(\theta) =0. 
\end{align*}
Consequently, it suffices to show that
\[\int_{\theta_L^*}^{\BAR\theta}\left[\nu^*(\theta)-\nu(\theta)\right]\left[\frac{\theta}{c} - \frac{\int_{\theta}^{\BAR\theta}\left[\a-\omega(s)\right]\ \dd F(s)}{\a cf(\theta)} - \psi(\nu^*(\theta))\right]\ \dd F(\theta)\geq 0.\]
However, this follows from our construction of $\nu_{0}$ in \cref{eq:nu_mu_positive} (cf.~\citealp{toikka11}).
\end{enumerate}

Following the above discussion, we conclude that  $(\und U^*,\nu^*)$ maximizes the Lagrangian, given the Lagrange multipliers $\mu^*$ and $\La^*$:
\[(\und U^*,\nu^*)\in\argmax_{\und U\in\R_+,\,\nu\in\calI}\calL(\und U,\nu;\mu^*,\La^*).\]
Note that the results of \cite{toikka11} imply that the $\nu^*$ can be alternatively be written as
\begin{equation}\label{eq:pos}
    \nu^*(\theta) = \psi^{-1}\(\BAR{\(s\mapsto \frac{s}{c}+\frac{\int_{\und\theta}^{s}\left[\a-\omega(z)\right]\ \dd F(z) -\La^*(s)}{\a cf(s)}\)}(\theta)\).
\end{equation}

%

\subsubsection*{\ul{Applying the Luenberger sufficiency theorem}.}

We now apply the \citeauthor{luenberger69} sufficiency theorem.  To this end, observe that:
\begin{enumerate}[label=({\bf\arabic*})]
\item {\bf $(\und U^*,\nu^*)$ satisfies the \eqref{eq:LS}  constraint.}

Since $\und U^*=\und U^{\LF}\geq \und\theta \nu^{\LF}(\und\theta) = \und\theta\nu^*(\und\theta)$, we conclude that the \eqref{eq:LS} constraint holds.

\item {\bf $(\und U^*,\nu^*)$ satisfies the \eqref{eq:IR}  constraint.}

Since $\und U^*=\und U^{\LF}$ and $\nu^*\geq \nu^{\LF}$, it follows from the envelope theorem (cf.~\Cref{clm:IC}) that the \eqref{eq:IR} constraint holds for all $\theta\in[\und\theta,\BAR\theta]$.

\item {\bf The complementary slackness conditions are satisfied.}

Since $\mu^*=0$, it is clear that $\mu^*\left[\und\theta\nu^*(\und\theta)-\und U^*\right] = 0$.

Moreover, observe that
\begin{align*}
\int_{\und\theta}^{\BAR\theta}\left[\und U^* - \und U^{\LF} + \int_{\und\theta}^{\theta}\left[\nu^*(s)-\nu^{\LF}(s)\right]\ \dd s \right]\ \dd\La^*(\theta)=\int_{\und\theta}^{\theta_L^*}\int_{\und\theta}^\theta\left[\nu^*(s)-\nu^{\LF}(s)\right]\ \dd s = 0.
\end{align*}
It follows that the desired complementary slackness condition holds.
\end{enumerate}

\subsubsection{Case \texorpdfstring{\#}{}2: \texorpdfstring{$\E[\omega]> \a$}{E[ω]>α}}

Finally, we consider the case where $\E[\omega] > \a$.  Instead of the social planner's problem, however, we study the following relaxation.
\begin{align*}
    \max_{\und U\in\R,\,\nu\in\calI}&\left\{\left[\E[\omega]-\a\right]\und U +\int_{\und\theta}^{\BAR\theta}\left[\left[\a\theta-\frac{\int_\theta^{\BAR\theta}\left[\a-\omega(s)\right]\ \dd F(s)}{f(\theta)}\right]\nu(\theta)-\a c \Psi(\nu(\theta))\right]\ \dd F(\theta)\right\}\\
    \text{s.t. }&\und U \leq \und\theta\nu(\und\theta).
\end{align*}
The solution to this relaxed problem coincides exactly with the solution to the full mechanism design problem studied in \Cref{sec:full_strong_analysis}:
\begin{equation}\label{eq:partial_strong_solution_inferior}
    \und U^* = \und\theta\nu^*(\und\theta)\AND\nu^*(\theta) = \psi^{-1}\(\BAR{\(s\mapsto \frac{J(s)}{c} + \frac{\E[\omega]-\a}{\a cf(s)}\cdot \und\theta\d_{\und\theta}(s)\)}(\theta)\).
\end{equation}
Since $\omega(\cdot)$ is increasing and $\E[\omega]-\a>0$, it follows that the distortion term is nonnegative for every consumer type.  Consequently, we have
\[\nu^*(\theta)\geq \nu^{\LF}(\theta)\qquad\forall\ \theta\in[\und\theta,\BAR\theta].\]
In particular, we see that $\und U^* = \und\theta\nu^*(\und\theta)\geq \und\theta\nu^{\LF}(\und\theta)\geq \und U^{\LF}$.
Moreover, for any $\theta\in[\und\theta,\BAR\theta]$, we must have
\[\int_{\und\theta}^\theta \nu^*(s)\ \dd s\geq \int_{\und\theta}^\theta \nu^{\LF}(s)\ \dd s.\]
It follows that, for any $\theta\in[\und\theta,\BAR\theta]$, the \eqref{eq:IR} constraint is satisfied:
\[\und U^* + \int_{\und\theta}^{\theta}\nu^*(s)\ \dd s\geq \und U^{\LF} + \int_{\und\theta}^\theta\nu^{\LF}(s)\ \dd s.\]

\subsection{Alternative Proof of \texorpdfstring{\Cref{thm:scope}}{Theorem 1}}

Although we provided a proof sketch of \Cref{thm:scope} in \Cref{sec:scope_proof}, we now provide an alternative proof using our Lagrangian approach.  Rather than explicitly construct a local positive deviation from the laissez-faire outcome, we prove \Cref{thm:scope} below by considering when the \eqref{eq:IR} constraints of all consumers bind.

As \cref{eq:partial_strong_solution,eq:partial_strong_solution_2,eq:partial_strong_solution_inferior} indicate, the social planner can always strictly improve on the laissez-faire outcome if $\E[\omega]>\a$, which implies that $\max\omega>\a$.  Below, we focus on the case where $\E[\omega]\leq\a$.

When welfare weight is negatively correlated with willingness to pay, the social planner can strictly improve on the laissez-faire outcome if and only if $\theta_H(\mu^*)>\und\theta$.  If $\mu^*>0$, then \cref{eq:mu,eq:La,eq:neg} imply that the social planner can strictly improve on the laissez-faire outcome; moreover, $\mu^*>0$ also implies that $\mu_{\max}>0$, which means that $\max\omega>\a$.  If $\mu^*=0$, then $\theta_H(\mu^*)>\und\theta$ implies there exists some $\hat\theta\in[\und\theta,\BAR\theta]$ such that
\[\int_{\und\theta}^{\hat\theta}\left[\a-\omega(s)\right]\ \dd F(s) >0.\]
Since $\theta\mapsto\int_{\und\theta}^{\theta}\left[\a-\omega(s)\right]\ \dd F(s)$ is quasiconvex and evaluates to zero at $\theta=\und\theta$, the condition above holds if $\max\omega>\a$.  Conversely, if $\max\omega>\a$, then either $\mu^*>0$ (in which case we obtain the desired result) or $\mu^*=0$; similar arguments as those above then show that $\theta_H(\mu^*)>\und\theta$.

When welfare weight is positively correlated with willingness to pay, the social planner can strictly improve on the laissez-faire outcome if and only if $\theta_L^*<\und\theta$.  Then there exists some $\hat\theta\in[\und\theta,\BAR\theta]$ such that
\[\int_{\hat\theta}^{\BAR\theta}\left[\a-\omega(s)\right]\ \dd F(s) <0.\]
Since $\theta\mapsto\int_\theta^{\BAR\theta}\left[\a-\omega(s)\right]\ \dd F(s)$ is quasiconvex and evaluates to zero at $\theta=\BAR\theta$, the condition above holds if $\max\omega>\a$.  This argument extends to the converse direction too.

\subsection{Proof of \texorpdfstring{\Cref{thm:optimal_negative}}{Theorem 2}}

If $\E[\omega]\leq\a$, then \cref{eq:mu,eq:La,eq:neg} provide the expression of the optimal allocation function $q^*$ in \Cref{thm:optimal_negative}.  If $\E[\omega]>\a$, then \cref{eq:partial_strong_solution,eq:partial_strong_solution_2} provide the expression of the optimal allocation function $q^*$ in\Cref{thm:optimal_negative}.

\subsection{Proof of \texorpdfstring{\Cref{thm:optimal_positive}}{Theorem 3}}

If $\E[\omega]\leq\a$, then \cref{eq:La_positive,eq:pos} provide the expression of the optimal allocation function $q^*$ in \Cref{thm:optimal_positive}.  If $\E[\omega]>\a$, then \cref{eq:partial_strong_solution_inferior} provides the expression of the optimal allocation function $q^*$ in \Cref{thm:optimal_positive}.

\clearpage

\renewcommand{\thesection}{Appendix B} 
\section{Additional Proofs}
\label{app:additional_proofs}
\renewcommand{\thesection}{B}
\setcounter{equation}{0}


\subsection{Proof of \texorpdfstring{\Cref{lem:active_negative_1}}{Lemma 1}}

Suppose that the private market is active, so there exists an interval of types $(\theta_-,\theta_+)$ such that $q^*(\theta)=q^{\LF}(\theta)$ for any $\theta\in(\theta_-,\theta_+)$.  Let $\kappa\in(\theta_-,\theta_+)$.  Similar to our approach \ref{app:proof_main}, we use \Cref{clm:IC,clm:LS} to rewrite the social planner's problem (given the optimal $\und U^*$) as follows:
\begin{align*}
    \max_{\nu\in\calI}&\int_{\kappa}^{\BAR\theta}\left[\left[\a\theta-\frac{\int_\theta^{\BAR\theta}\left[\a-\omega(s)\right]\ \dd F(s)}{f(\theta)}\right]\nu(\theta)-\a c \Psi(\nu(\theta))\right]\ \dd F(\theta)\\
    \text{s.t. }&
        \int_{\kappa}^\theta \nu(s)\ \dd s \geq \int_{\kappa}^{\theta} \nu^{\LF} (s)\ \dd s\qquad\text{for any }\theta\in[\kappa,\BAR\theta].
\end{align*}
We proceed by considering the necessary conditions that the optimal subutility function $\nu^*$ must satisfy.  To this end, we let $\La$ denote the Lagrange multiplier for the \eqref{eq:IR} constraint.  We can then write the Lagrangian as
\[\calL(\nu;\La) = \int_{\kappa}^{\BAR\theta}\left[\left[\a\theta+\frac{\La(\theta)-\int_\theta^{\BAR\theta}\left[\a-\omega(s)\right]\ \dd F(s)}{f(\theta)}\right]\nu(\theta)-\a c \Psi(\nu(\theta))+\La(\theta)\nu^{\LF}(\theta)\right]\ \dd F(\theta).\]
The necessary conditions for the Lagrange approach \citep{luenberger69} ensure that there exists a nonincreasing function $\La^*:[\und\theta,\BAR\theta]\to\R_+$ such that 
\[\psi(\nu^*(\theta)) = \BAR{\left.\(s\mapsto \frac{s}{c} + \frac{\La^*(s) - \int_s^{\BAR\theta}\left[\a-\omega(z)\right]\ \dd F(z)}{\a cf(s)}\)\right|_{[\kappa,\BAR\theta]}}(\theta)\qquad\text{for any }\theta\in[\kappa,\BAR\theta].\]
Moreover, since the \eqref{eq:IR} constraint binds for $(\kappa,\theta_+)$, observe that for any $\theta\in(\kappa,\theta_+)$,
\[0 = \psi(\nu^{\LF}(\theta)) - \psi(\nu^{*}(\theta)) = \frac{\La^*(\theta)-\int_\theta^{\BAR\theta}\left[\a-\omega(s)\right]\ \dd F(s)}{\a cf(\theta)}\implies \La^*(\theta) =\int_\theta^{\BAR\theta}\left[\a-\omega(s)\right]\ \dd F(s).\]
Since $\theta\mapsto\int_\theta^{\BAR\theta}\left[\a-\omega(s)\right]\ \dd F(s)$ is a quasiconcave function when $\omega$ is decreasing, the \eqref{eq:IR} constraint can bind only for types $\theta$ such that $\omega(\theta)\leq\a$ (\ie, sufficiently high types).

Now, suppose that there is a ``gap'' between two open subintervals of $[\kappa,\BAR\theta]$ where the \eqref{eq:IR} constraint binds.  Denote this ``gap'' by $(\kappa_-,\kappa_+)$.  
To show that such a ``gap'' cannot happen, it suffices to show that the following condition holds for any nondecreasing subutility function $\nu\in\calI$ such that $\int_{\kappa_-}^{\theta}\nu(s)\ \dd s\geq \int_{\kappa_-}^\theta\nu^{\LF}(s)\ \dd s$ for any $\theta\in(\kappa_-,\kappa_+)$, and $\int_{\kappa_-}^{\kappa_+}\nu(s)\ \dd s= \int_{\kappa_-}^{\kappa_+}\nu^{\LF}(s)\ \dd s$:
\[\int_{\kappa_-}^{\kappa_+}\left[\nu^{\LF}(\theta)-\nu(\theta)\right]\left[\frac{J(\theta)}{c}-\frac{\theta}{c}\right]\ \dd F(\theta)\geq 0.\]
This is the variational inequality that verifies that $\nu^{\LF}$ is optimal on this interval.  But this variational inequality is equivalent to
\begin{align*}
&\int_{\kappa_-}^{\kappa_+} \left[\nu(\theta)-\nu^{\LF}(\theta)\right]\int_{\theta}^{\BAR\theta}\left[\a-\omega(s)\right]\ \dd F(s)\ \dd\theta \geq0\\
&\iff \int_{\kappa_-}^{\kappa_+} \underbrace{\left[\a-\omega(\theta)\right]}_{\geq 0}\underbrace{\int_{\kappa_-}^{\theta}\left[\nu(s)-\nu^{\LF}(s)\right]\ \dd s}_{\geq 0}\ \dd F(\theta)\geq0.
\end{align*}
Clearly, the latter inequality holds.  Consequently, there can be no ``gaps'' in the region where the \eqref{eq:IR} constraint binds: if the \eqref{eq:IR} constraint binds over a region in $[\kappa,\BAR\theta]$, then this region must be an interval.

Finally, suppose that the \eqref{eq:IR} constraint binds over the interval $(\kappa_-,\kappa_+)$, but $\kappa_+<\BAR\theta$.  Then we repeat the above argument on the interval $(\kappa_+,\BAR\theta)$: it suffices to show that
\begin{align*}
&\int_{\kappa_+}^{\BAR\theta}\left[\nu^{\LF}(\theta)-\nu(\theta)\right]\left[\frac{J(\theta)}{c}-\frac{\theta}{c}\right]\ \dd F(\theta)\geq 0\\
&\iff \int_{\kappa_+}^{\BAR\theta} \left[\nu(\theta)-\nu^{\LF}(\theta)\right]\int_{\theta}^{\BAR\theta}\left[\a-\omega(s)\right]\ \dd F(s)\ \dd\theta \geq0\\
&\iff \int_{\kappa_+}^{\BAR\theta} \underbrace{\left[\a-\omega(\theta)\right]}_{\geq 0}\underbrace{\int_{\kappa_+}^{\theta}\left[\nu(s)-\nu^{\LF}(s)\right]\ \dd s}_{\geq 0}\ \dd F(\theta)\geq0.
\end{align*}
So, if the \eqref{eq:IR} constraint binds, it must bind on an interval in $[\kappa,\BAR\theta]$ that ends at the highest possible type $\BAR\theta$.  Thus, if the \eqref{eq:IR} constraint binds for some type $\kappa\in[\und\theta,\BAR\theta]$, it must also bind for $\BAR\theta$.

\subsection{Proof of \texorpdfstring{\Cref{lem:active_negative_2}}{Lemma 2}}

Our proof of \Cref{lem:active_negative_1} did not address whether or when the \eqref{eq:IR} constraint binds in $[\und\theta,\kappa]$.  We now analyze when the \eqref{eq:IR} constraint binds in $[\und\theta,\BAR\theta]$ by using \Cref{lem:active_negative_1}: that we can, without loss of generality, take $\kappa=\BAR\theta$ to be where the \eqref{eq:IR} constraint binds.  

We modify the social planner's problem by first rewriting the \eqref{eq:LS} constraint.  Observe that the utility of the lowest type, $\und U$, can be expressed as
\[\und U = \BAR U^{\LF} - \int_{\und\theta}^{\BAR\theta}\nu(s)\ \dd s.\]
Thus the \eqref{eq:LS} constraint can be equivalently written as
\[\und U \leq \und\theta\nu(\und\theta)\iff \int_{\und\theta}^{\BAR\theta}\nu(s)\ \dd s + \und\theta\nu(\und\theta)\geq \BAR U^{\LF}.\]
We therefore write the social planner's problem as follows:
\begin{align*}
\max_{\nu\in\calI}&\int_{\und\theta}^{\BAR\theta}\Bigg[\underbrace{\left[\a\theta + \frac{\int_{\und\theta}^\theta\left[\a-\omega(s)\right]\ \dd F(s)}{f(\theta)}\right]}_{=:\a H(\theta)}\nu(\theta) - \a c\Psi(\nu(\theta))\Bigg]\ \dd F(\theta)\\
\text{s.t. }&\begin{dcases}\int_{\und\theta}^{\BAR\theta}\nu(s)\ \dd s + \und\theta\nu(\und\theta)\geq \BAR U^{\LF},\\\int_{\theta}^{\BAR\theta} \nu(s)\ \dd s\leq \int_{\theta}^{\BAR\theta} \nu^{\LF}(s)\ \dd s.\end{dcases}
\end{align*}
Denote the Lagrange multiplier function for the \eqref{eq:IR} constraint by $\La:[\und\theta,\BAR\theta]\to\R_+$, which is a nondecreasing function satisfying $\La(\und\theta)=0$.  Let $\mu\in\R_+$ be the Lagrange multiplier for the \eqref{eq:LS} constraint.  The Lagrangian can thus be written as
\[\calL(\nu;\La,\mu) = \int_{\und\theta}^{\BAR\theta}\left[\left[H(\theta)+\frac{\mu + \mu\und\theta\cdot\d_{\und\theta}(\theta) - \La(\theta)}{\a f(\theta)}\right]\nu(\theta) - c\Psi(\nu(\theta))\right]\ \dd F(\theta).\]

As before, we use necessary conditions to pin down what the optimal Lagrange multiplier function $\La^*$ must satisfy.  First, observe that the necessary conditions for the Lagrange approach \citep{luenberger69} ensures that there exists a nondecreasing function $\La^*:[\und\theta,\BAR\theta]\to\R_+$ such that 
\[\psi(\nu^*(\theta)) = \BAR{\(s\mapsto\frac{H(s)}{c}+\frac{\mu^*+\mu^*\und\theta\cdot\d_{\und\theta}(s) - \La^*(s)}{\a cf(s)}\)}(\theta)\qquad\forall\ \theta\in[\und\theta,\BAR\theta].\]
Moreover, if the \eqref{eq:IR} constraint binds for an interval $(\theta_-,\theta_+)$, then 
\[\psi(\nu^*(\theta))=\psi(\nu^{\LF}(\theta)) = \frac{H(\theta)}{c} + \frac{\mu^*+\mu^*\und\theta\cdot\d_{\und\theta}(\theta) - \Lambda^*(\theta)}{\a cf(\theta)}.\] 
This implies that
\begin{equation}\label{eq:Lagrange_multiplier_app}
\Lambda^*(\theta) = \left[\frac{H(\theta)}{c}-\psi(\nu^{\LF}(\theta))\right]\a cf(\theta) + \mu^* + \mu^*\und\theta\d_{\theta=\und\theta}\qquad\forall\ \theta\in(\theta_-,\theta_+).
\end{equation}
Observe that
\[\left[\frac{H(\theta)}{c}-\psi(\nu^{\LF}(\theta))\right]\a cf(\theta) = \int_{\und\theta}^{\theta}\left[\a-\omega(s)\right]\ \dd F(s),\]
which is a quasiconvex function in $\theta$ since $\frac{\pp}{\pp\theta}\int_{\und\theta}^{\theta}\left[\a-\omega(s)\right]\ \dd F(s) = \left[\a-\omega(\theta)\right]f(\theta)$, which crosses zero at most once since $\omega$ is nonincreasing.
Moreover, the quasiconvexity of the right-hand side of \eqref{eq:Lagrange_multiplier_app} is preserved when we add $\mu^* + \mu^*\und\theta\d_{\und\theta}$.  Therefore, the \eqref{eq:IR} constraint can bind only for sufficiently high types (\ie, when $\omega(\theta)\leq \a$).

Now, suppose that there is a ``gap'' between two open intervals where the \eqref{eq:IR} constraint binds.  Denote this ``gap'' by $(\kappa_-,\kappa_+)$.  
To show that such a ``gap'' cannot happen, it suffices to show that for any $\nu\in\calI$ such that $\int_{\kappa_-}^{\theta}\nu(s)\ \dd s\geq \int_{\kappa_-}^\theta\nu^{\LF}(s)\ \dd s$ for any $\theta\in(\kappa_-,\kappa_+)$, and $\int_{\kappa_-}^{\kappa_+}\nu(s)\ \dd s= \int_{\kappa_-}^{\kappa_+}\nu^{\LF}(s)\ \dd s$,
\[\int_{\kappa_-}^{\kappa_+}\left[\nu^{\LF}(\theta)-\nu(\theta)\right]\left[\frac{H(\theta)}{c}-\frac{\theta}{c}\right]\ \dd F(\theta)\geq 0.\]
This is the variational inequality that verifies that $\nu^{\LF}$ is optimal on this interval.  But this variational inequality is equivalent to
\begin{align*}
&\int_{\kappa_-}^{\kappa_+} \left[\nu^{\LF}(\theta)-\nu(\theta)\right]\int_{\und\theta}^{\theta}\left[\a-\omega(s)\right]\ \dd F(s)\ \dd\theta \geq0\\
&\iff \int_{\kappa_-}^{\kappa_+} \underbrace{\left[\a-\omega(\theta)\right]}_{\geq 0}\underbrace{\int_{\kappa_-}^{\theta}\left[\nu(s)-\nu^{\LF}(s)\right]\ \dd s}_{\geq 0}\ \dd F(\theta)\geq0.
\end{align*}
Clearly, the latter inequality holds.  Consequently, there can be no ``gaps'' in the region where the \eqref{eq:IR} constraint binds: if the \eqref{eq:IR} constraint binds over a region in $[\und\theta,\BAR\theta]$, then this region must be an interval.

\subsection{Proof of \texorpdfstring{\Cref{prop:benefit_negative}}{Proposition 1}}

\Cref{prop:benefit_negative} follows from applying the envelope theorem (cf.~\Cref{clm:IC}) to the characterization of the optimal allocation function given by \Cref{thm:optimal_negative}.

\subsection{Proof of \texorpdfstring{\Cref{prop:benefit_positive}}{Proposition 2}}

\Cref{prop:benefit_positive} follows from applying the envelope theorem (cf.~\Cref{clm:IC}) to the characterization of the optimal allocation function given by \Cref{thm:optimal_positive}.

\subsection{Proof of \texorpdfstring{\Cref{prop:non-market}}{Proposition 3}}

Given that $\und\theta>0$ by assumption, the optimal mechanism includes a free public option if and only if $\mu^*>0$.  The result of \Cref{prop:non-market} then follow from the characterizations of $\mu^*$ given by \Cref{thm:optimal_negative,thm:optimal_positive}.

\subsection{Proof of \texorpdfstring{\Cref{prop:comp_statics}}{Proposition 4}}

\begin{enumerate}[label={\em(\roman*)}]
\item When $\omega$ is decreasing, we rewrite $\mu^*$ as
\[\min\left\{\mu\in[\(\E[\omega]-\a\)_+,\mu_{\max}]:\int_{\und\theta}^{\BAR\theta}\left[v(q^{\LF}(s))-v(q_\mu(s))\right]\ \dd s\leq \und\theta\left[v(q_\mu(\und\theta))-v(q^{\LF}(\und\theta))\right]\right\}.\]
Observe that $\mu\mapsto\int_{\und\theta}^{\BAR\theta}\left[v(q^{\LF}(s))-v(q_\mu(s))\right]\ \dd s$ is pointwise decreasing in $\a$ while $\mu\mapsto\und\theta\left[v(q_\mu(\und\theta))-v(q^{\LF}(\und\theta))\right]$ is pointwise increasing in $\a$.  Since the former is a decreasing function of $\mu$ and the latter is an increasing function of $\mu$, we conclude that $\mu^*$ must decrease with $\a$.  Moreover, $\theta_H$ is a decreasing function of $\mu$; hence $\theta_H(\mu^*)$ increases with $\a$.

\item When $\omega$ is increasing, observe that $\theta_L^*$ decreases with $\a$ since $\theta\mapsto\int_{\theta}^{\BAR\theta}\left[\a-\omega(s)\right]\ \dd F(s)$ is pointwise decreasing in $\a$.  Moreover, $\mu^*=\(\E[\omega]-\a\)_+$ also decreases with $\a$.
\end{enumerate}

\subsection{Proof of \texorpdfstring{\Cref{prop:topping}}{Proposition 5}}

\begin{enumerate}[label={\em(\roman*)}]
\item When $\omega$ is decreasing, it suffices to show that the optimal allocation characterized in \Cref{thm:optimal_negative} is always distorted downwards for types just below $\theta_H(\mu^*)$, so the social planner must strictly benefit from preventing topping up whenever there is scope for in-kind redistribution, as characterized by \Cref{thm:scope}.  On one hand, if $\E[\omega]\leq\a$, then we consider two cases: either $\mu^*=0$ or $\mu^*>0$.  If $\mu^*=0$, then it is easy to see from \cref{eq:neg} that the distortion for types just below $\theta_H(\mu^*)$ is negative.  If $\mu^*>0$, then the distortion for those types can fail to be negative only if the distortion for {\em all} types in $(\und\theta,\theta_H(\mu^*))$ is positive; but this is impossible since $\mu^*>0$ and \cref{eq:mu} together imply that
\[\und\theta \nu^*(\und\theta) + \int_{\und\theta}^{\theta_H^*} \nu^*(s)\ \dd s = U^{\LF}(\theta_H^*) = \und U^{\LF} + \int_{\und\theta}^{\theta_H^*}\nu^{\LF}(s)\ \dd s.\]
On the other hand, if $\E[\omega]>\a$, then a similar argument applies using \cref{eq:partial_strong_solution_2} instead.  

\item When $\omega$ is increasing, it suffices to show that the optimal allocation characterized in \Cref{thm:optimal_positive} satisfy the stronger \eqref{eq:IR'} constraints.  This is easy to see from \cref{eq:pos,eq:partial_strong_solution_inferior}, as we discuss in \ref{app:proof_main} following those equations.
\end{enumerate}

\end{document}